\documentclass[11pt,a4paper]{article}
\usepackage{lmodern}
\pdfoutput=1
\usepackage{jheppub}
\usepackage{rotating}
\usepackage{array}
\usepackage{amsmath}
\usepackage[normalem]{ulem}
\usepackage{slashed}
\usepackage{booktabs}
\usepackage[pdftex,table,usenames]{xcolor}
\usepackage{enumerate}
\usepackage{parskip}
\usepackage{upgreek}
\usepackage{layouts}
\usepackage{appendix}
\usepackage{siunitx}
\usepackage{paralist}
\usepackage{multirow}
\usepackage{xspace}
\usepackage{tikz}
\usepackage{hyperref}
\usepackage{mathrsfs}
\usepackage{pgfplots}
\usepackage{fontawesome5}
\usepackage{silence}
\usepackage{footmisc}
\usepackage{wrapfig}
\usepackage{xspace}
\usepackage{rotating}
\usepackage{newtxtext} 
\usepackage{amssymb}
\usepackage{graphicx}
\usepackage{siunitx}
\usepackage{epstopdf}
\usepackage{epsfig}
\usepackage[numbers,sort&compress]{natbib}
\usepackage{color}
\usepackage{url}
\usepackage{eurosym}
\usepackage{color,colortbl,listings}
\usepackage{array}
\usepackage{multirow}
\usepackage{lscape}
\usepackage{booktabs}
\usepackage{xspace}
\usepackage{rotating}
\usepackage[utf8x]{inputenc}
\usepackage[section]{placeins}
\usepackage{eurosym}
\usepackage{multirow}
\usepackage{pgfgantt}
\usepackage{xcolor}
\usepackage{tcolorbox}
\usepackage{subcaption} 
\usepackage{gensymb}

\DeclareSIUnit\barn{b}

\newcommand{\exclude}[1]{}

\arxivnumber{xxxx.xxxxx}
\preprint{MPP-2026-209}

\title{Search for dark matter axions at 1\text{--}2 \SI{}{\micro\electronvolt} with a RADES-DarkQuantum haloscope at CERN}

\author[a]{Alberto Arcusa Puente}
\author[l]{Sergio Arguedas Cuendis}
\author[c]{Marcos Buetas Mata}
\author[b]{Sergio Calatroni}
\author[d]{Cristian Cogollos}
\author[e]{Alejandro Díaz-Morcillo}
\author[d]{Babette D\"obrich}
\author[f]{Jacob Egge}
\author[d]{Elisa Gabbrielli}
\author[a]{Javier Galindo Guarch}
\author[g]{Jose María García-Barceló}
\author[h]{Benito Gimeno}
\author[g]{Yikun Gu}
\author[g]{Sophia Hollick}
\author[g]{Igor Garcia Irastorza}
\author[d]{David Kittlinger}
\author[h]{Adrian Lambies-Asensio}
\author[f]{Axel Lindner}
\author[i]{Manuel Meyer}
\author[e]{Adrián Muñoz-Urán}
\author[e]{Jose R. Navarro-Madrid}
\author[d]{Tobias Ortmann}
\author[c]{Sagar Paul}
\author[g]{Angel Rodriguez}
\author[j]{Kristof Schmieden}
\author[g]{Laura Segui}
\author[c,k]{Wolfgang Wernsdorfer}
\author[g]{Mina Yami}

\affiliation[a]{Instituto Tecnológico de Aragón (ITA), Maria de Luna 7-8, 50018 Zaragoza, Spain}
\affiliation[b]{CERN - European Organization for Nuclear Research, Geneva, Switzerland}
\affiliation[c]{Physikalisches Institut (PHI), Karlsruhe Institute of Technology (KIT), Wolfgang-Gaede-Str. 1, D-76131 Karlsruhe, Germany}
\affiliation[d]{Max-Planck-Institut f\"{u}r Physik (Werner-Heisenberg-Institut), Boltzmannstraße 8, 85748 Garching, Germany}
\affiliation[e]{Universidad Politécnica de Cartagena, Campus Muralla del Mar, C/ Doctor Fleming, s/n, 30202 Cartagena, Spain.}
\affiliation[f]{Deutsches Elektronen-Synchrotron DESY, Notkestr.\,85, 22607 Hamburg, Germany}
\affiliation[g]{Centro de Astropartículas y Física de Altas Energías (CAPA), Universidad de Zaragoza, 50009 Zaragoza, Spain}
\affiliation[h]{IFIC Valencia, Carrer del Catedrátic José Beltrán Martinez, 2, 46980 Paterna, Valencia, Spain}
\affiliation[i]{CP3-Origins, University of Southern Denmark (SDU), Campusvej 55, 5230 Odense M, Denmark}
\affiliation[j]{Rheinische Friedrich-Wilhelms-University, Nußallee 12, Bonn, Germany}
\affiliation[k]{Institute of Quantum Materials and Technologies (IQMT), Karlsruhe Institute of Technology (KIT), Hermann-von-Helmholtz-Platz 1, 76344, Eggenstein-Leopoldshafen, Germany}
\affiliation[l]{Consejo Nacional de Rectores (CONARE), San Jose, Costa Rica}

\emailAdd{rades-eb@cern.ch}

\abstract{
The QCD axion remains one of the most compelling candidates for the dark matter of the Universe, yet the mass range around $1$\text{--}\SI{2}{\micro\electronvolt} remains essentially unexplored. 
In this paper, we describe a proposal to employ CERN's M1 superconducting magnet 
to deploy 
a
haloscope to perform a search for galactic axion and axion-like particle dark matter providing continuous coverage from approximately 1.2 to 1.8 \SI{}{\micro\electronvolt}. 
Under conservative assumptions, the experiment can reach the QCD axion band over a significant fraction of the scanned range, while optimistic
detector performance would allow sensitivity to benchmark models such as KSVZ and DFSZ over a large part of the explored mass interval.
With the current long shut-down of CERN's accelerators providing a unique and timely window, 
such an experiment will deliver high-impact results even in conservative scenarios and potentially delivering a definitive test of the axion dark matter paradigm in the unexplored $1$\text{--}\SI{2}{\micro\electronvolt} mass range.}

\keywords{axion, haloscope, CERN, Dark Matter}

\begin{document}

\maketitle

\hyphenation{Baby-IAXO}

\section{Introduction}
\label{sec:introduction}

Axions are among the most compelling candidates for physics beyond the Standard Model (SM): The Peccei-Quinn (PQ) mechanism provides a natural solution to the strong-CP problem~\cite{Peccei:1977hh,Peccei:1977ur}, predicting a light, pseudoscalar particle—the axion—whose existence would suppress CP violation in QCD~\cite{Weinberg:1977ma,Wilczek:1977pj}. Beyond this, axions are a leading dark matter (DM) candidate: non-thermal production via vacuum realignment and topological defect decay can yield relic abundances consistent with observations, particularly in the $10^{-6} - 10^{-3}$ eV mass range~\cite{Tegmark:2005dy,Kawasaki:2014sqa,Ringwald:2015dsf}. This makes axions an extremely well-motivated particle and the number of experimental searches for it has been steeply increasing over the past years, see ~\cite{Irastorza:2015geo,DiLuzio:2021qct,Dobrich:2025oso} for  reviews.

\begin{figure}[ht]
 \centering
 \vspace{-1mm}
  \includegraphics[width=.55\textwidth]{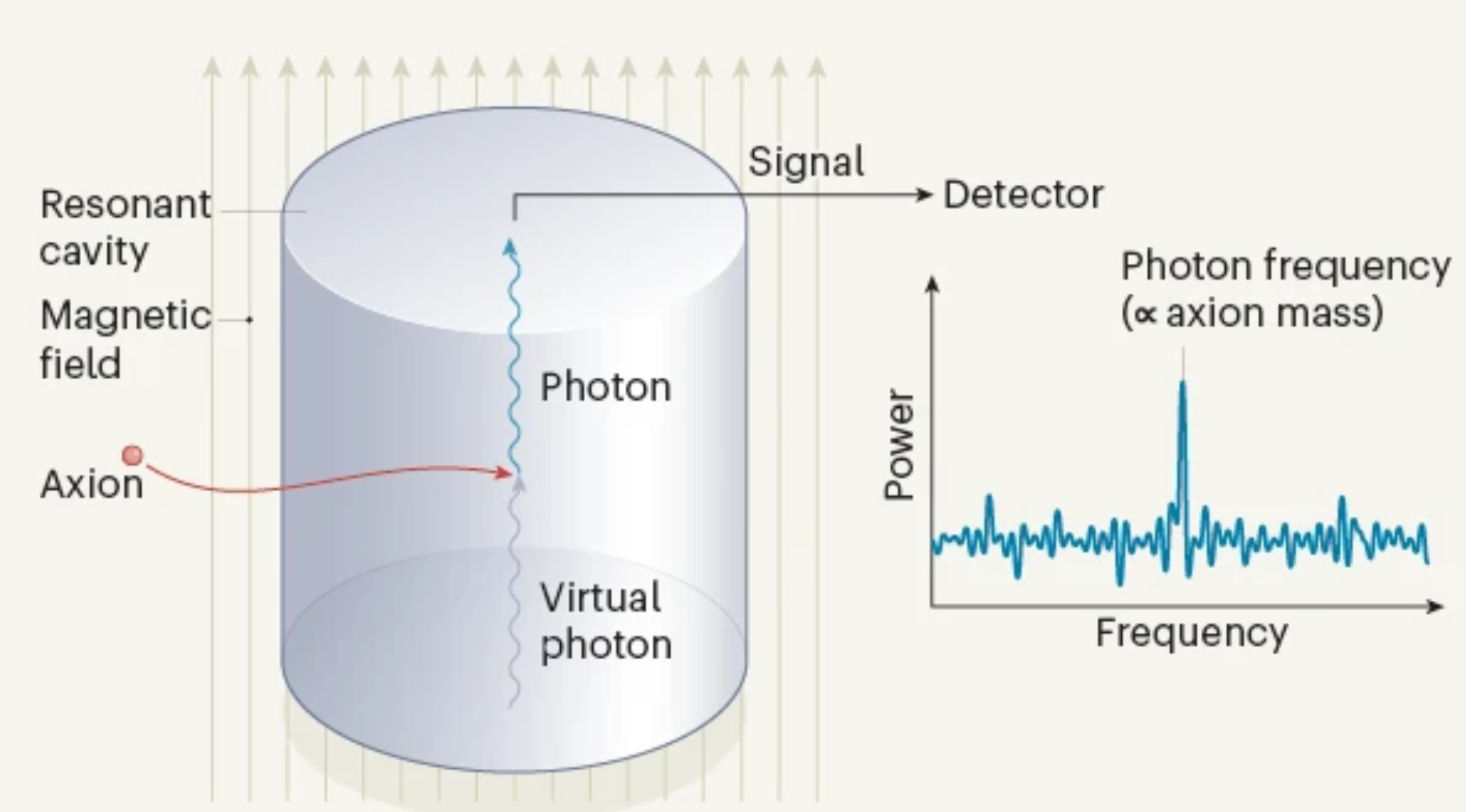}
  \caption{\footnotesize 
  Sketch of Sikivie's axion haloscope concept, from \cite{IrastorzaNature2021}
  .}
  \label{fig:haloscope}
 \end{figure}

The most established experimental approach to detect the DM axion is the \textit{haloscope}, based on Sikivie’s concept~\cite{Sikivie:1983ip}, sketched in Fig.~\ref{fig:haloscope}.
In a strong magnetic field, axions of mass $m_a$ convert to microwave photons at frequency $f_{\rm resonance}\propto m_a$,  enhanced by high-quality-factor ($Q$) cavities. A figure of merit in such experiments is, using the axion DM density $\rho_a$ and coupling $g_{a\gamma}$
\begin{equation}
F_{\rm halo} \propto \rho_a^2 \, g_{a\gamma}^4 \, m_a^2 \, B_0^4 \, V^2 \, T_{\rm sys}^{-2} \, Q .
\end{equation}
It shows the experimental need for large cavity volume $V$, high $Q$, low noise temperature $T_{\rm sys}$, and strong external magnetic field $B_0$. This is a challenge especially at low masses due to the requirement of large cavities resonating at low frequencies and the availability of large magnet bores.

\begin{figure}[t]
     \centering
     \includegraphics[width=0.7\linewidth]{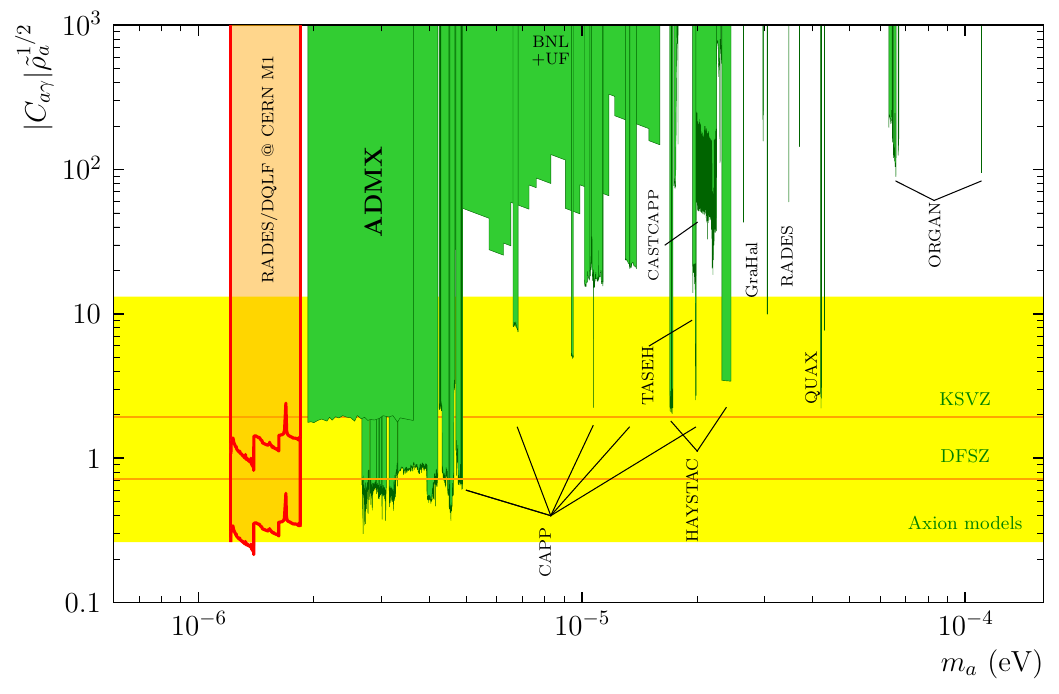}
\caption{
Current experimental constraints and projected sensitivity in the axion-photon parameter space, expressed in terms of the dimensionless axion-photon coupling $C_{a\gamma}\propto g_{a\gamma}/m_a$. The green regions show existing experimental exclusions, while the yellow band indicates the parameter space of benchmark QCD axion models, including the KSVZ and DFSZ benchmark lines. The orange regions show the projected sensitivity of the proposed RADES-DQLF search at the CERN M1 magnet for the optimistic (lower curve) and conservative (upper curve) scenarios. The assumptions entering these projections, including the cavity quality factor, physical temperature, receiver-added noise, coupling configuration and total exposure, are discussed in detail in Sec.~\ref{sec:sensitivity}. The horizontal extent of the DQLF projections indicates the mass range covered by the three proposed cavities, approximately 1.2\text{--}1.8 \SI{}{\micro\electronvolt}.
}
\label{fig:overview}
 \end{figure}

Fig.~\ref{fig:overview} shows the current status of affairs in DM axion searches: green are excluded areas from haloscope experiments while the yellow band indicated target region for `vanilla axion models' KSVZ~\cite{Kim:1979if,Shifman:1979if}  and DFSZ~\cite{Dine:1981rt,Zhitnitsky:1980tq}. 
While the ADMX experiment has excluded axions near 2\text{--}\SI{4}{\micro\electronvolt} and aims to reach $\sim$ \SI{10}{\micro\electronvolt}~\cite{ADMX:2018gho,ADMX:2006kgb}, the mass region just below ADMX’s reach ($\sim\,1$\text{--}\SI{2}{\micro\electronvolt}) remains unexplored—not due to theoretical or cosmological reasons, but because of the above-mentioned experimental difficulty. 

Indeed, although recently several proposals have been put forward (e.g., GraHal \cite{Pugnat:2024sxb}, (Princeton Axion Search) PXS \cite{Otto:2026fly,10858383} or FLASH~\cite{Alesini:2023qed},  exploiting potential access to existing large-scale magnet infrastructure), the mass range immediately below the ADMX lower edge remains completely unexplored experimentally. 

Another proposal to tackle this mass region was put forward by the RADES (Relic Axion Detector Exploratory Setup) collaboration. Based on experience with haloscope searches at the CERN \cite{CAST:2020rlf,Ahyoune:2024klt}, it has proposed to exploit the future BabyIAXO magnet—a 10-meter-long, 70-cm-diameter superconducting dipole with $B_0 \sim 2.5~\mathrm{T}$ and a total magnetic volume exceeding $7~\mathrm{m}^3$~\cite{Ahyoune:2023gfw}. The idea is the deployment of multiple meter-scale cavities, covering the 1\text{--}2 \SI{}{\micro\electronvolt} window with sensitivity to axion benchmark models.

On the other hand, BabyIAXO is still several years from operation.
To bridge this gap in time, we therefore propose to exploit a different window of opportunity:
With the current third long shut-down (LS3) of CERN's accelerator complex, the option exists to search for axions around 1\text{--}2 \SI{}{\micro\electronvolt} with CERN's M1 magnet, a 1.5-meter-bore, approximately 2-T superconducting dipole, located in the North Area. Due to the shut-down of the accelerators, the M1 magnet cannot be used in conjunction with beams during LS3 and is potentially available to physics purposes not requiring a beam. While smaller than BabyIAXO, M1 offers a large enough bore to host cavities of meter-scale, exactly what is needed to target the region immediately below ADMX.

In this proposal, we present the RADES-DQLF (Dark Quantum Low Frequency) experiment: a cryogenic, quantum-limited haloscope to be deployed in the CERN M1 magnet. Using three interchangeable 1-m-long cavities with different cross-sections, we propose to scan a sizeable parameter range in the 1\text{--}2 \SI{}{\micro\electronvolt} area. The experiment leverages the experience of the RADES collaboration in superconducting coatings~\cite{Golm:2021ooj} and  a SQUID-based readout, developed under the ERC Synergy Grant \textit{DarkQuantum} \cite{cordisDQ}. This setup will probe previously unexplored parameter space with competitive sensitivity, independently of a future RADES search in BabyIAXO, as sketched in Fig.~\ref{fig:overview}.

Crucially, the RADES-DQLF experiment is not just a prototype. It is a complete, scientifically compelling axion search. It will validate key technologies—cryogenics, superconducting coatings, quantum-limited readout—for the future BabyIAXO implementation, while delivering early physics results in a compelling mass window.

\section{Experimental setup}

The proposed RADES-DQLF experiment follows the standard axion haloscope concept described in section~\ref{sec:introduction}. 
 The cavity resonance is tuned across the target axion-mass range, while the resulting microwave signal is extracted and amplified by a low-noise receiver. In the present implementation, three frequency-tunable resonant cavities will be operated sequentially inside the CERN M1 magnet and coupled to a cryogenic, near-quantum-limited SQUID-based readout. The experimental design builds on the cavity concepts and technologies developed within RADES and DarkQuantum, adapted to the specific magnetic, cryogenic and mechanical constraints of M1. The following sections describe the main components of the experiment, including the resonant cavities and their tuning system, the development of superconducting coatings to enhance their quality factor, the receiver chain, comprising the SQUID-based first-stage amplifier, the subsequent amplification and data-acquisition chain at room temperature, and finally the cryogenic system required for operation at millikelvin temperatures.

\subsection{Resonant cavities and tuning mechanism}

\subsubsection{Resonant cavity design concept}
\label{subsec:cavity_design}

\begin{figure}[ht]
    \centering
    \includegraphics[width=0.9\linewidth]{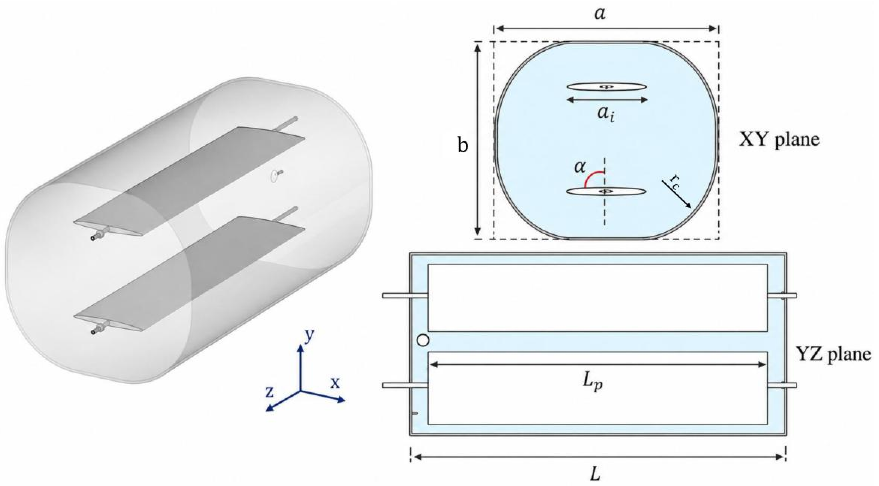}
    \caption{ \footnotesize  Schematic view of the cavities geometry. Left: 3D view of the cavity showing the tuning plates. In the 3D view, only one port can be shown. Right: XY and YZ plane cuts of the cavity showing the dimensions. The angle $\alpha$ of the plates is defined with respect to the $x$-axis. Tuning rods that hold the plates in place within the cavity have a radius of 5 mm for all the DQLF cavity designs.}
    \label{fig:prototype_view}
\end{figure}

The cavity designs proposed are based on a quasi-cylindrical geometry derived from the RADES-BabyIAXO cavity concept. According to the study in \cite{Ahyoune:2023gfw}, the resonant mode that provides the maximum form factor in this type of cavities is the quasi-$\mathrm{TE}_{111}$ $y$-polarization (named $\mathrm{qTE}_{111-\mathrm{y}}$), which is therefore selected as the axion or operation mode. The first cavity (called DQLF-1) has the same cross-sectional dimensions as the largest RADES-BabyIAXO cavity. The second and third cavities (DQLF-2 and -3) have smaller transversal dimensions, in order to cover a higher frequency range. A schematic view of the proposed cavities geometry is depicted in Fig. \ref{fig:prototype_view}, while the dimensions of the three designs and their corresponding tuning ranges are summarised in Table \ref{tab:Cav_dimensions}. 

\begin{table}[ht]
\centering
\begin{tabular}{| c | c | c | c | c | c|}
\hline
 Cavity model & $a$ (cm) & $b$ (cm) & $a_i$ (cm) & $r_c$ (cm) &Freq. range  \\
\hline
DQLF-1 & 56.0 & 50.4 & 15.0 & 22.0 & 293.88 - 337.02 MHz  \\ \hline
DQLF-2 & 45.8 & 41.2 & 13.0 & 17.6 & 336.07 - 391.76 MHz \\ \hline
DQLF-3 & 39.2 & 35.2 & 10.5 & 13.6 & 388.69 - 445.63 MHz \\

\hline
\end{tabular}
\caption{\footnotesize  Dimensions and operational frequency range of the three cavities. The cavity length ($L=1$ m) and the plates length ($L_p=89$ cm) is the same for all models. The thickness of both metallic plates is 1 cm.}
\label{tab:Cav_dimensions}
\end{table}
\newpage
Similar to the original cavities, the tuning mechanism is based on two rotating metal plates of rectangular geometry aligned with the longitudinal axis and placed inside the resonant volume. As they rotate, the electric field distribution within the cavity is modified, thereby changing the resonant frequency of some modes. Following a study into the optimal geometry of the plates and after the measurements comparison with two types of geometry, it was concluded that an oval cross-section should be used, as this allowed for better flow of the surface currents generated and, consequently, a smaller reduction in the quality factor (see Fig. \ref{fig:prototype_view}).

The cavities include two coaxial loop ports: Port 1 is used to detect the electromagnetic signal generated by the axion, whereas Port 2 is used to characterise the cavity and will remain short-circuited during data acquisition. Both ports are based on a coaxial loop that allows the signal to be extracted via coupling with the magnetic field. Consequently, Port 1 is situated at the centre of one of the cavity end-caps, where the magnetic field of the $\mathrm{qTE}_{111-\mathrm{y}}$ mode is at its maximum. 

To ensure a good RF contact between the plates axes, ports and the cavity walls, a solution based on finger gaskets surrounding the axes is proposed. These components, made of beryllium copper, have already been used by groups within the collaboration \cite{garcíabarceló2026vortexcavityradesaxion}. Thanks to their flexibility, they allow to keep the electrical contact, therefore reducing the possible radiation produced by these parts.

\begin{figure}[h!]
    \centering
    \includegraphics[angle=90, width=0.5\textwidth]{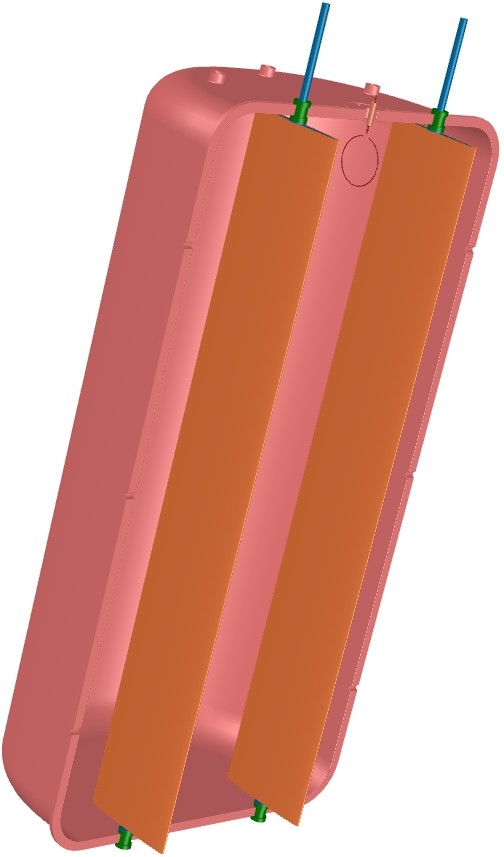}
    \caption{3D cross-sectional rendering of the DQLF-1 cavity.}
    \label{fig:3D-DQLF-1}
\end{figure}

A complete engineering design of the cavities has been carried out by the CERN engineering support group EN-MME, in view of their fabrication at CERN. Current baseline is to fabricate the cavities out of 316L stainless steel plates, minimizing thickness taking the advantage that no net pressure is exerted  over the cavity body. The cavities will then be plated  with a galvanic copper coating following either the "shiny" or "matt" procedures described in~\cite{Golm:2021ooj}. This choice results from a compromise of having reasonable mechanical stability during field ramping of the magnet, which prevents the use of solid copper due to eddy currents and the consequent mechanical forces, and adequate RF quality factor and thermal conductivity at cryogenic temperatures. To facilitate deposition, the geometrical radius between the cavity outer wall and the endcaps 
has been set to 220~mm for DQLF-1. These cavities will optionally receive an NbTi superconducting coating, which will be described in more detail in section~\ref{sec:superconducting_coating}. A 3D longitudinal section rendering of the DQLF-1 cavity is shown in Fig.~\ref{fig:3D-DQLF-1}, DQLF-2 and DQLF-3 being essentially identical except for the smaller dimensions.

\subsubsection{Cavities characterisation: Form factor and quality factor} 
\label{sec:cav_characterisation}

\subsubsection*{Electromagnetic simulations of the cavities}

Electromagnetic simulations were carried out for the three proposed DQLF cavities in order to evaluate their tuning range, form factor (obtained from \ref{eq:C}), unloaded quality factor and cavity figure of merit ($\mathrm{FoM}=Q_0V^2C^2$). In all cases, the axion mode ($\mathrm{qTE}_{111-\mathrm{y}}$) was tracked. The cavity walls and tuning plates are assumed to have the conductivity of $1\cdot 10^9$ for OFHC copper at the expected cryogenic operating temperature. For the form factor calculation, the simulated M1 magnetic field data, provided by CERN, have been used (see section \ref{sec:layout_fringe_fields}).

\begin{equation}
\label{eq:C}
    C =
    \frac{\left|\int_V \vec{E}\cdot\vec{B}_0\,dV\right|^2}
    {\int_V |\vec{B}_0|^2\,dV
    \int_V \varepsilon_r |\vec{E}|^2\,dV},
\end{equation}

\noindent where $\vec{E}$ is the electric field of the axion-sensitive mode, $\vec{B}_0$ is the external magnetic field, and $\varepsilon_r$ is the relative permittivity inside the cavity, which is equal to unity in vacuum.

\begin{figure}[h]
\centering
\begin{subfigure}[b]{0.47\textwidth}
         \centering
         \includegraphics[width=1\textwidth]{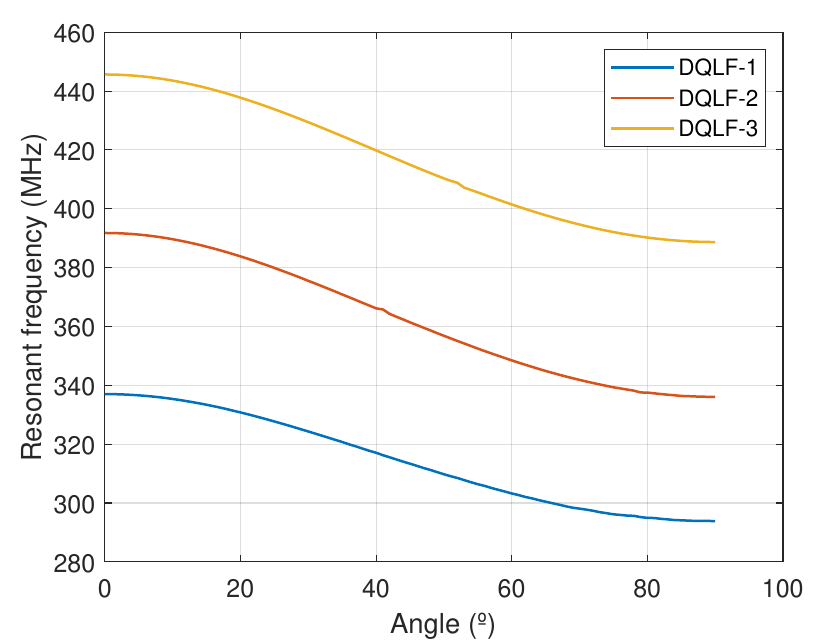}
         \caption{}
         \label{fig:Frequency}
\end{subfigure}
\hfill
\begin{subfigure}[b]{0.47\textwidth}
         \centering
         \includegraphics[width=1\textwidth]{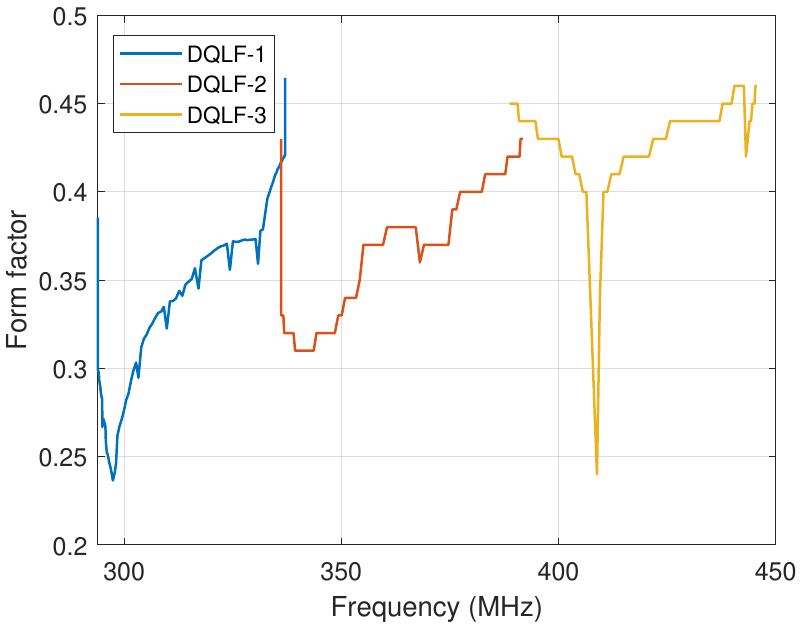}
         \caption{}
         \label{fig:FormFactor}
\end{subfigure}
\hfill
\begin{subfigure}[b]{0.47\textwidth}
         \centering
         \includegraphics[width=1\textwidth]{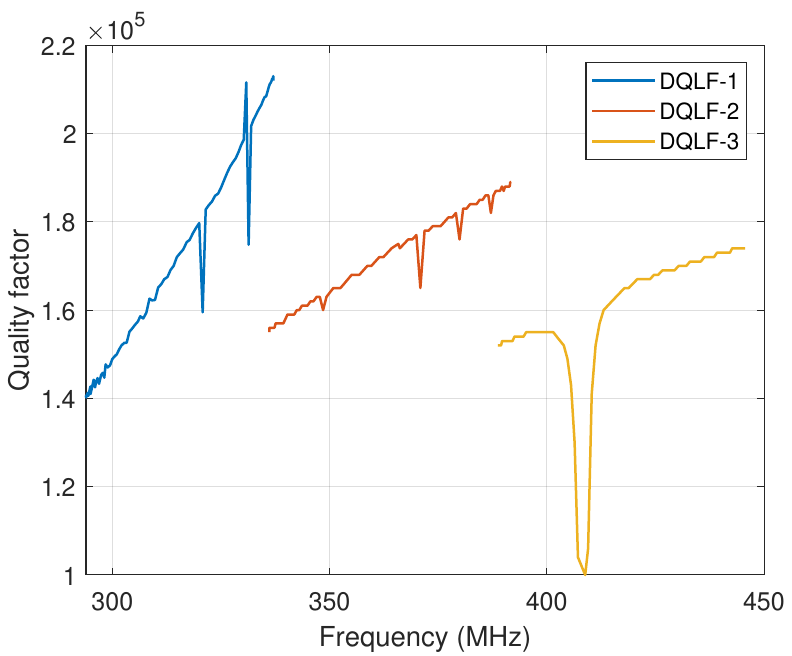}
         \caption{}
         \label{fig:UnloadedQ}
\end{subfigure}
\hfill
\begin{subfigure}[b]{0.47\textwidth}
         \centering
         \includegraphics[width=1\textwidth]{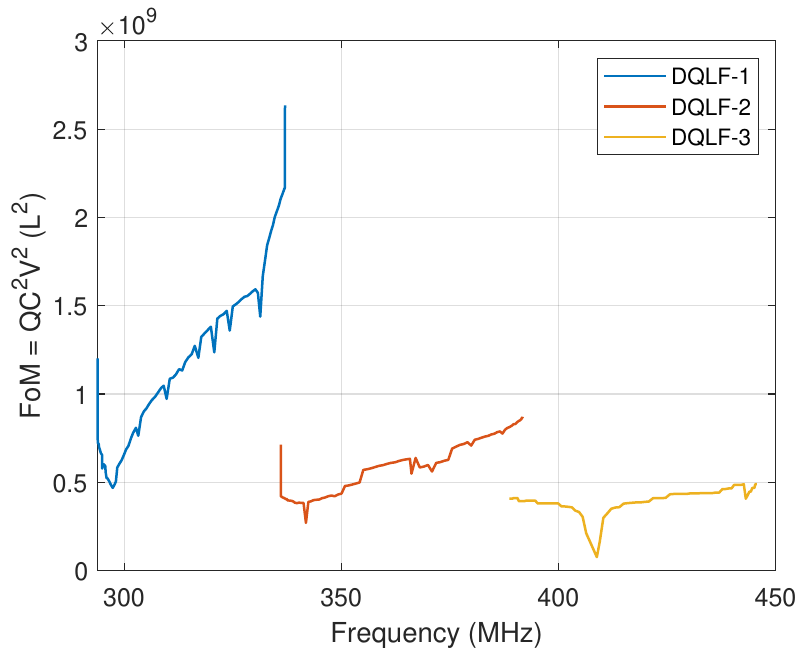}
         \caption{}
         \label{fig:FoM}
\end{subfigure}
\caption{
Electromagnetic simulation results for the three DQLF cavity geometries,
DQLF-1, DQLF-2 and DQLF-3, tracking the axion mode $\mathrm{qTE}_{111-\mathrm{y}}$
throughout the tuning range. (a) Resonant frequency as a function of the
tuning-plate angle, showing how the three cavities provide continuous
frequency coverage from approximately 294 to 446~MHz, with overlapping
tuning ranges between adjacent cavities. (b) Axion form factor $C$,
(c) unloaded quality factor $Q_0$, and (d) resulting cavity figure of
merit as functions of resonant frequency. Local variations and sharp
features arise from the evolution of the resonant mode during tuning,
including mode crossings. The simulations assume the baseline
opposite-rotation configuration of the two tuning plates.
}
\label{fig:cavity_characterisation}
\end{figure}

A relevant point studied during the design phase was the relative rotation direction of the two tuning plates. Two possible configurations were considered: rotating both plates in the same direction or rotating them in opposite directions. The simulations showed that the opposite rotation of the plates provides better electromagnetic performance, both in terms of form factor and unloaded quality factor, with improvements of up to a 13\% and 11\%, respectively. This behaviour is attributed to a more favourable deformation of the field pattern during the tuning process, which reduces the degradation of the mode structure and of the surface current distribution on the metallic walls and plates. Therefore, the opposite turn configuration was selected as the baseline tuning strategy for the cavities. The simulated results for all three designs are shown in Fig.~\ref{fig:cavity_characterisation}.

Fig.~\ref{fig:cavity_characterisation}a shows the resonant frequency as a function of the tuning-plate angle. The combination of the three cavities provides continuous frequency coverage over the full target range, with overlapping regions around $337~\mathrm{MHz}$ and $390~\mathrm{MHz}$. As shown in Fig.~\ref{fig:cavity_characterisation}b, the form factor of DQLF-1 remains mostly within $C \simeq 0.25$--$0.45$, with local variations associated with the tuning process. DQLF-2 and DQLF-3 exhibit values of the same order, typically within $C \simeq 0.31$--$0.43$ and $C \simeq 0.24$--$0.46$, respectively. These values confirm that the axion mode maintains a substantial overlap with the external magnetic field throughout most of the tuning range.

Fig.~\ref{fig:cavity_characterisation}c shows the unloaded quality factor $Q_0$. For DQLF-1, $Q_0$ increases from approximately $1.4\cdot10^5$ to values above $2.1\cdot10^5$ across the tuning range. DQLF-2 exhibits values between approximately $1.55\cdot10^5$ and $1.9\cdot10^5$, while DQLF-3 ranges from approximately $1.0\cdot10^5$ to $1.74\cdot10^5$. The observed variations are mainly related to the redistribution of surface currents as the tuning plates rotate. Local drops are also present, most likely associated with mode crossings or distortions of the resonant field pattern.

Finally, Fig.~\ref{fig:cavity_characterisation}d shows the resulting cavity figure of merit. 
DQLF-1 reaches the highest values due to a larger volume. DQLF-2 and DQLF-3 exhibit lower, but still significant, FoM values and extend the sensitivity to higher frequencies, thus completing the continuous coverage of the target frequency range.

\subsubsection*{Measurements of the DQLF-1 cavity}
\label{sec:kit_meas}
\begin{figure}[ht]
\centering
 \includegraphics[height=0.5\textwidth]{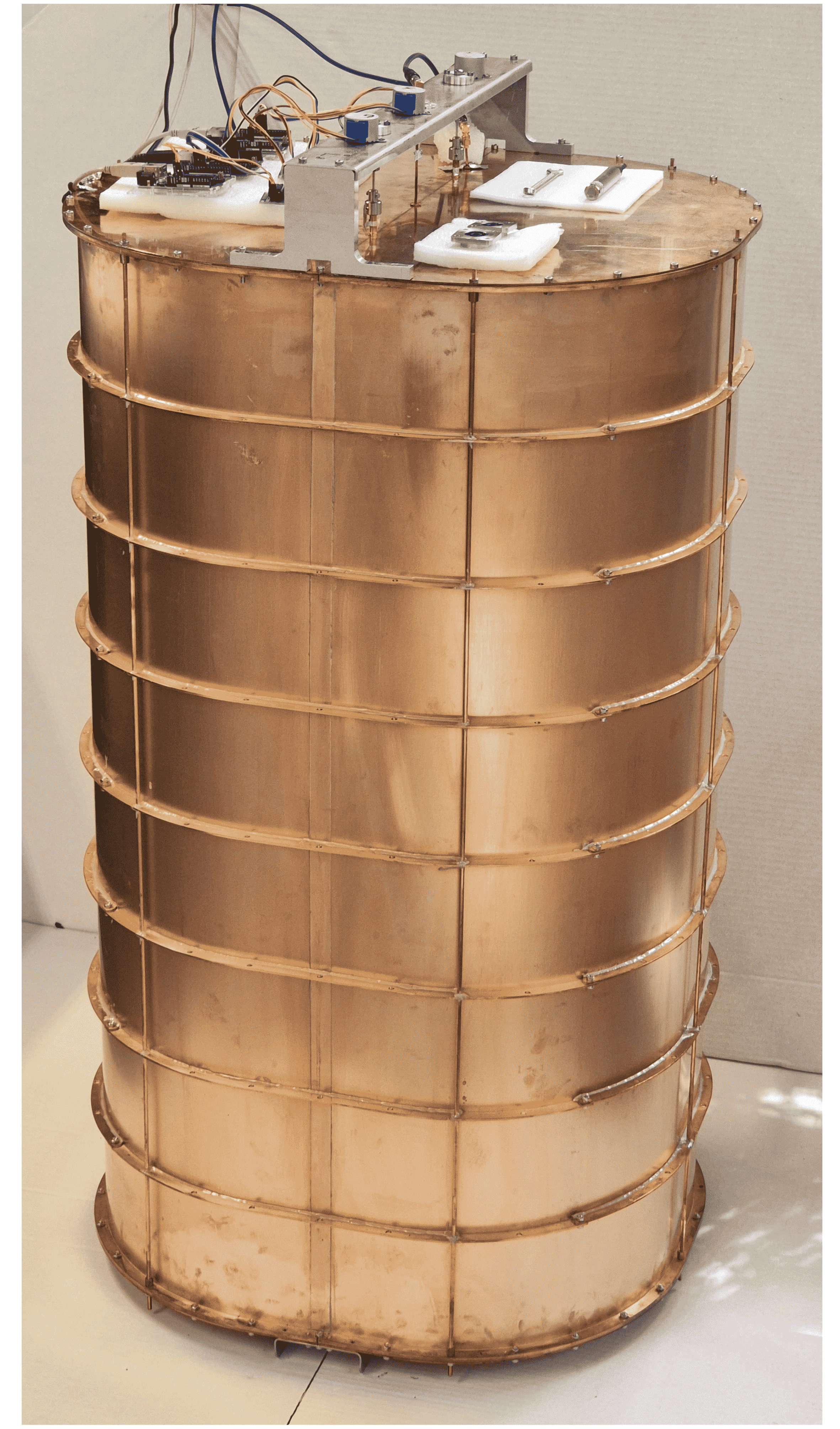} \hspace{2mm}
 \includegraphics[height=0.5\textwidth, trim={0cm 0 0cm 0},clip]{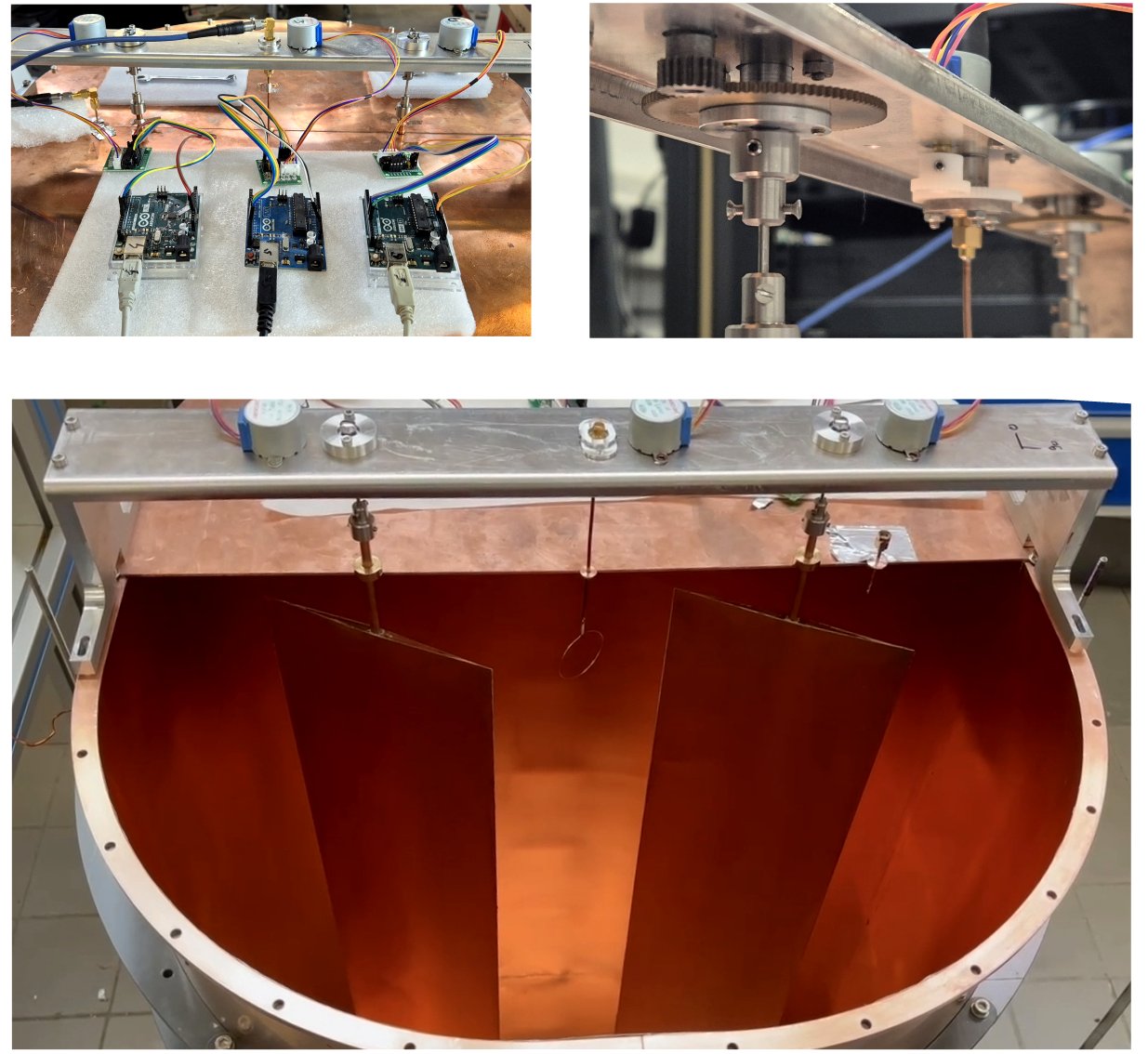}
 \caption{ Left: The DQLF-1 cavity with external copper rings and rods implemented for mechanical support and thermalization. 
Top right: Enlarged view of the motors and gears for the tuning plates and antenna rotation.
Bottom right: Inner view of the cavity showing the tuning plates with triangular cross-section used during the preliminary measurements.
}
\label{fig:kit_cavity}
\end{figure}

The built first version of the DQLF-1 cavity is shown in Fig.~\ref{fig:kit_cavity}, left. The cavity walls were constructed from OFHC copper sheets, mechanically supported by a system of external ribs that provided structural rigidity and maintained the required quasi-cylindrical geometry. A support bar was installed along the top of the cavity, carrying the motors used to rotate the coupling loop and the tuning plates. In this first implementation, the tuning plates had a triangular cross-section formed by two joined metal sheets (see Fig.~\ref{fig:kit_cavity}, right).

At this stage, measurements were performed at each motor step ($0.703^\circ$), covering a total rotation of $90^\circ$ with 128 measurement points. Initial tests of the automatic re-coupling software, which was being developed in parallel, were also carried out. The coupling between the readout port and the resonant mode was found to be relatively weak in this first configuration, most likely due to the dimensions of the coupling loop, providing useful input for the optimization of the final readout implementation.

\begin{figure}[ht]
     \centering
     \includegraphics[width=0.65\linewidth]{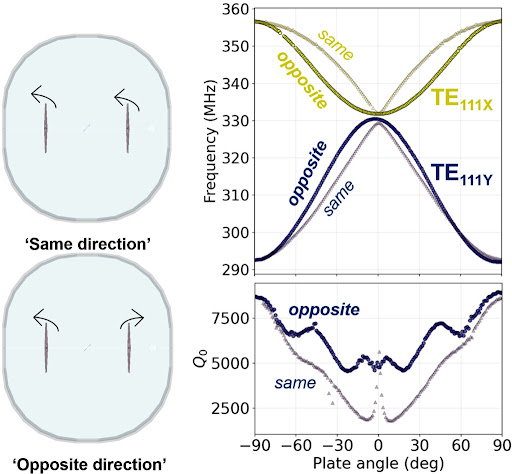}
     \caption{
Characterization of the DQLF-1 tuning system at room temperature.
Left: Cross-section of the cavity illustrating the two tuning configurations investigated, corresponding to rotation of the two tuning plates in the same or in opposite directions.
Right: Measured resonant frequency (top) and unloaded quality factor $Q_0$ (bottom) as a function of the tuning-plate angle for the two configurations. In the upper-right panel, two resonant modes, $\mathrm{qTE}_{111-\mathrm{y}}$ and $\mathrm{qTE}_{111-\mathrm{x}}$, are shown; only the $\mathrm{qTE}_{111-\mathrm{y}}$ mode has a significant axion form factor and is therefore relevant for the axion search. The measurements demonstrate the expected frequency tuning over the full angular range and confirm the dependence of the cavity quality factor on the relative direction of rotation of the tuning plates, supporting the choice of the opposite-rotation configuration for the final design.
}
     \label{fig:Measurements_KIT}
 \end{figure}

Two complete data sets were acquired to characterize the cavity and experimentally compare the two tuning configurations, with the plates rotating either in the same or in opposite directions (see Fig.~\ref{fig:Measurements_KIT}). The measurements confirmed the behavior anticipated from the electromagnetic simulations, in particular the dependence of the quality factor on the relative direction of rotation of the tuning plates. Small differences in the resonant frequency were also observed for some plate angles. These are not relevant for the final operation, since the tuning motors provide sufficient precision to position the cavity at the desired frequency.

The measurements also provided important input for the mechanical optimization of the cavity. The measured quality factor was below the value predicted by the room-temperature simulations, and dedicated studies identified the construction of the tuning plates as the dominant source of additional RF losses. In the first prototype, each plate was assembled from two metal sheets joined to form a triangular cross-section, introducing a longitudinal discontinuity in the conducting surface and consequently perturbing the RF surface-current flow. This observation motivated the adoption of a smooth oval cross-section for the final tuning-plate design. Measurements with the improved geometry demonstrated an approximately $100\%$ increase in $Q_0$ with respect to the original triangular configuration, validating the revised design for the subsequent DQLF implementation.

\subsubsection{Tuning and coupling systems}

\begin{figure}[ht]
    \centering
    \includegraphics[width=0.7\linewidth]{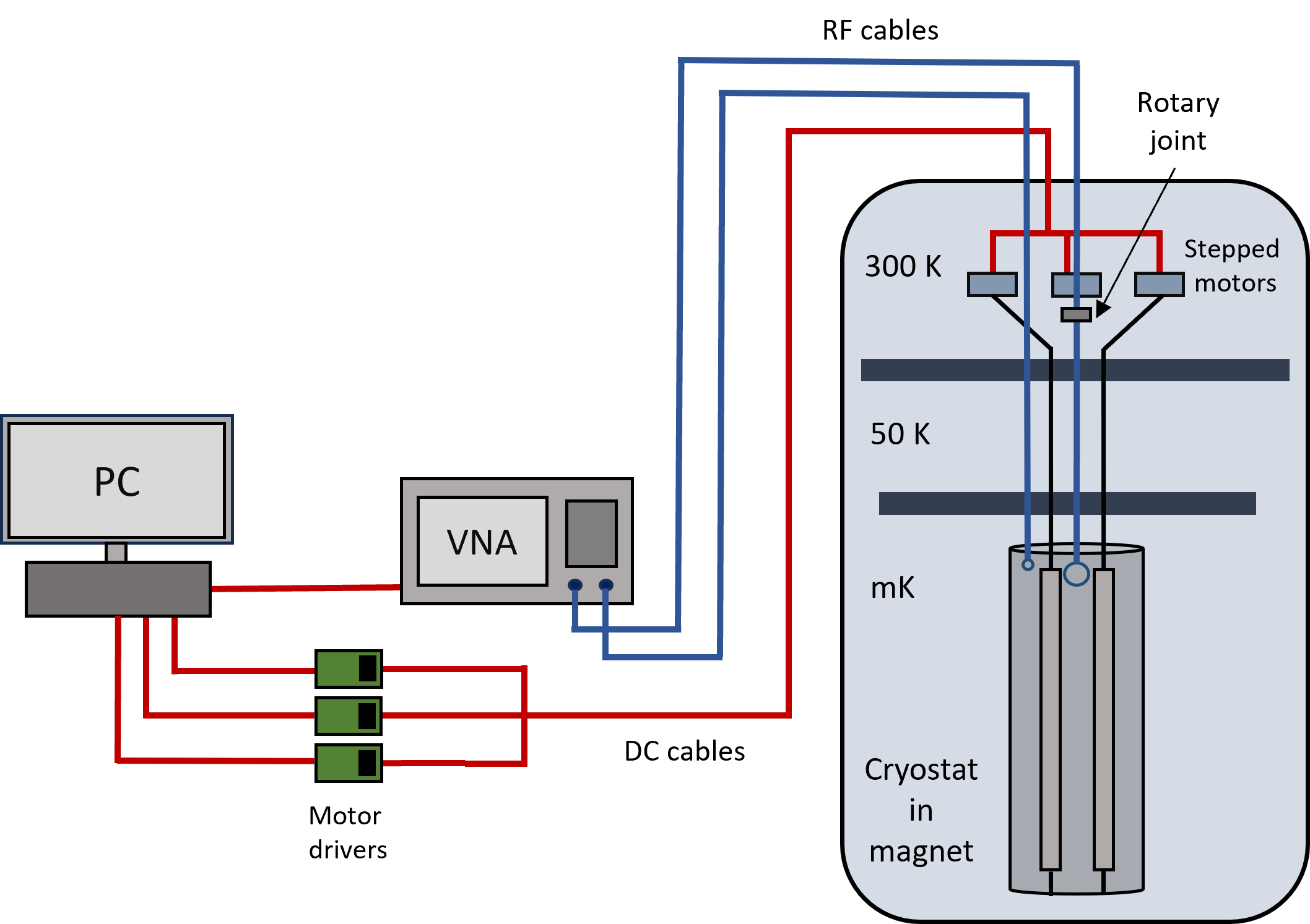}
    \caption{Scheme of the tuning system setup for the DQLF cavities, based on the controlled rotation of two internal metallic plates.}
    \label{fig:setup_tuning}
\end{figure}

The cavities to be operated in the M1 magnet will require to simultaneously and precisely move both rotation plates. Figure \ref{fig:setup_tuning} shows the tuning concept for the cavity experiment. 

A key requirement of the tuning system is the ability to position the cavity resonance with a frequency resolution significantly finer than its loaded bandwidth. This provides sufficiently precise control of the resonance frequency for reproducible positioning and for implementing the desired spacing between consecutive data-taking points. For the expected operating conditions, the relevant frequency scale is set by the loaded quality factor, $Q_{\mathrm{l}}$, and hence by the $-3\,\mathrm{dB}$ bandwidth of the resonance. For a cavity coupled through a single readout port,
\begin{equation}
    Q_{\mathrm{l}} = \frac{Q_0}{1+\beta},
\end{equation}
where $\beta$ is the port coupling coefficient. For DQLF-1 with cryogenic-copper surfaces and an expected electrical conductivity of $1\cdot10^9$ S/m, the electromagnetic simulations predict a maximum unloaded quality factor of approximately $Q_0 = 2 \cdot 10^5$. Assuming $\beta=2$, this corresponds to $Q_{\mathrm{l}} \simeq 6.7 \cdot 10^4$. At a representative resonant frequency of $337\,\mathrm{MHz}$, the corresponding $-3\,\mathrm{dB}$ bandwidth is
\begin{equation}
    \Delta f_{-3\,\mathrm{dB}} = \frac{f_r}{Q_{\mathrm{l}}}
    \simeq 5.05\,\mathrm{kHz}.
\end{equation}
As a design requirement, we require the minimum frequency-setting increment of the tuning mechanism to be approximately one order of magnitude smaller than the resonance bandwidth. This corresponds to a tuning resolution of about $0.5\,\mathrm{kHz}$ for the cryogenic-copper cavity. This minimum tuning increment should not be confused with the spacing between consecutive data-taking points in the axion scan, which is determined by the adopted scan strategy and is discussed in section~\ref{sec:sensitivity}.

The use of superconducting surfaces, in particular an NbTi coating as discussed in section~\ref{sec:superconducting_coating}, would increase the cavity quality factor and consequently impose a more demanding requirement on the tuning mechanism. Assuming $Q_0 \simeq 2 \cdot 10^6$, the corresponding minimum frequency-setting increment would be of order $0.05\,\mathrm{kHz}$.

The required frequency resolution can be translated into a mechanical angular-resolution requirement using the simulated dependence of the resonant frequency on the tuning-plate angle. For a small angular displacement,
\begin{equation}
    \delta f \simeq \frac{df}{d\alpha}\,\delta\alpha,
\end{equation}

where $df/d\alpha$ is obtained from the electromagnetic simulations of the cavity. Requiring the frequency increment to remain below one tenth of the resonance bandwidth therefore gives
\begin{equation}
    \delta\alpha_{\rm max}
    = \min_{\alpha}\left[
        \left|\frac{d\alpha}{df}\right|
        \frac{\Delta f_{-3\,\mathrm{dB}}}{10}
    \right].
\end{equation}

Using the frequency-step criterion described above, the minimum angular step is estimated to be approximately \SI{90}{\micro\relax}${}^\circ$ for the cryogenic copper and \SI{9}{\micro\relax}${}^\circ$ for the NbTi case. These values define the target resolution for the plate actuation system and impose strong requirements for the motor characteristics, the mechanical transmission, backlash control, and reproducibility after successive tuning steps.

During the frequency scan, the rotation of the tuning plates may also modify the electromagnetic coupling between the cavity mode and the readout port. As a consequence, $\beta$ will vary as a function of the plates angle. Since a stable and well-defined coupling condition is needed for the measurements, the tuning system must be complemented by an automatic re-coupling procedure. The control software will therefore include a feedback routine able to acquire the $S$-parameters from the VNA, extract the resonant frequency, $Q_{\mathrm{l}}$, $Q_0$, and $\beta$, and adjust the angular position of the coupling loop when the measured coupling deviates from the target value. This re-coupling algorithm will be an essential element during the data-taking campaign, ensuring that each frequency point is measured under controlled coupling conditions.

The measurement sequence will follow an automated tuning-coupling logic. Starting from a defined initial plates angle, the system will acquire and store the $S$-parameters, determine the resonance parameters, verify whether the desired coupling has been achieved, and then move the plates to the next frequency point. If the target coupling is not reached, the port loop angle will be varied before continuing the scan. For operation inside the M1 magnet, the motorization system must be compatible with the experimental environment. The motors must operate under vacuum and room temperature conditions, provide sufficient torque to rotate the internal tuning plates, offer the angular precision required by the frequency-resolution criterion, and be compatible with the magnetic field environment of the magnet. Previous measurements with earlier motors showed limitations such as backlash and non-smooth rotation, which motivates the use of a more robust piezoelectric solution for the experiment.

The selected actuator for the tuning and coupling system is the \textit{SMARACT SR-9219C-S-HV-NM-C120-P02} rotation stage \cite{SMARACT_motors}. This model has been chosen as the baseline solution for the M1 measurements. The use of piezoelectric motors is particularly suitable for this application because it allows precise angular positioning, compact integration, and operation under the constraints imposed by the M1 magnet setup. The final implementation will include two positioners for the rotation of the tuning plates and an additional actuator for the port loop rotation, allowing the frequency scan and re-coupling procedure to be performed in a fully controlled and automated way.

\subsubsection{Optimal coupling}

\label{sec:optimal_coupling}

As stated above, a key figure of merit of a haloscope experiment is the scanning rate, which quantifies how rapidly a given mass range can be explored at a specified sensitivity. The scanning rate depends, among other parameters, on the cavity-to-receiver coupling coefficient $\beta$, which can therefore be optimised for a given unloaded cavity quality factor $Q_0$, axion quality factor $Q_a\approx10^6$, and receiver-noise configuration. We follow here the formalism of Ref.~\cite{CAPP_revisiting}.

For a receiver chain including a circulator terminated by a matched load, the optimal coupling is given by
\begin{equation}
    \beta_{\mathrm{opt}}=
    \frac{1}{2}\left(1+\sqrt{9+8Q_0/Q_a}\right).
\end{equation}
In the absence of a circulator, the impedance mismatch between the cavity and the receiver must also be taken into account, and $\beta_{\mathrm{opt}}$ is obtained by solving Eq.~(5.2) of Ref.~\cite{CAPP_revisiting}. In this case, the optimum also depends on the relative contributions of the cavity and receiver noise.

Table~\ref{tab:beta_opt} provides an illustrative example of the coupling values obtained from this optimisation for the three proposed cavities. For this comparison, we consider a representative cavity temperature $T_{\mathrm{cav}}=100~\mathrm{mK}$ and receiver-added noise temperature $T_{\mathrm{add}}=50~\mathrm{mK}$, corresponding to $\lambda_T=T_{\mathrm{add}}/T_{\mathrm{cav}}=0.5$. Results are shown both for the reference cryogenic-copper cavity and for a superconducting configuration in which $Q_0$ is assumed to be ten times higher, and for receiver configurations with and without a circulator. The values in the table are intended to illustrate the dependence of the optimal coupling on the cavity quality factor and receiver configuration, rather than to define the operating conditions adopted for the sensitivity projections. The latter are discussed in detail in section~\ref{sec:sensitivity}, where $\beta_{\mathrm{opt}}$ is evaluated consistently with the parameters of each sensitivity scenario.

\begin{table}[ht]
    \centering
    \begin{tabular}{|l|c|c|c|c|}
        \hline
        \multirow{2}{*}{Cavity}
        & \multicolumn{2}{c|}{With circulator}
        & \multicolumn{2}{c|}{Without circulator} \\
        \cline{2-5}

        & $\beta_{\mathrm{opt}}$ (Cu)
        & $\beta_{\mathrm{opt}}$ (NbTi)
        & $\beta_{\mathrm{opt}}$ (Cu)
        & $\beta_{\mathrm{opt}}$ (NbTi) \\
        \hline

        DQLF-1 & 2.09--2.13 & 2.75--3.05 & 8.6--8.7 & 10.5--11.4 \\
        \hline
        DQLF-2 & 2.05--2.12 & 2.81--2.95 & 8.6--8.7 & 10.7--11.2 \\
        \hline
        DQLF-3 & 2.09--2.13 & 2.56--2.89 & 8.6--8.7 & 10.7--11.0 \\
        \hline
    \end{tabular}
    \caption{Illustrative optimal coupling coefficients for the three DQLF cavities, considering cryogenic-copper (Cu) and superconducting (NbTi) surfaces and receiver configurations with and without a circulator. The values are calculated for the representative noise parameters $T_{\mathrm{cav}}=100~\mathrm{mK}$ and $T_{\mathrm{add}}=50~\mathrm{mK}$ and are intended to illustrate the dependence of $\beta_{\mathrm{opt}}$ on the cavity and receiver configuration. The coupling values used for the sensitivity scenarios are discussed in section~\ref{sec:sensitivity}.}
    \label{tab:beta_opt}
\end{table}

The results illustrate the different trade-offs involved in the optimization. In particular, in the absence of a circulator, the formal scan-rate optimum can occur at relatively large values of $\beta$. Stronger overcoupling reduces the loaded quality factor and hence broadens the cavity bandwidth, while simultaneously modifying the fraction of the axion signal delivered to the receiver and the cavity-noise contribution. The optimum therefore results from the balance of these effects rather than from maximising the signal power at a single cavity setting. In practice, the accessible coupling range will also be constrained by the geometry of the coupling loop and by the mechanical range and reproducibility of the automatic re-coupling system. These experimental constraints, together with the noise assumptions appropriate to each configuration, are incorporated in the sensitivity scenarios discussed in section~\ref{sec:sensitivity}.

\subsection{Superconducting coating option} 
\label{sec:superconducting_coating}

The baseline fabrication approach for the DQLF cavities is to manufacture the body of the cavities with stainless steel AISI~316L, coated with an electroplated copper layer of a minimum thickness of \SI{10}{\micro\meter}, as discussed in \ref{subsec:cavity_design}. To further enhance performance beyond the limits of a copper coating alone, it is planned to develop the option of depositing an NbTi layer on the cavity, on top of the existing copper layer, with the aim of achieving at least a tenfold increase in the cavity $Q_0$. NbTi coatings have been already discussed and applied within the QUAX experiment, see~\cite{Carugno2019,PhysRevD.99.101101,Braggio2024,PhysRevX.15.021031}. These coatings have been developed and tested at frequencies in the range from 7 to 14~GHz, and demonstrated to achieve quality factors at least $\times 10$ better than copper at the same operating conditions and at an applied magnetic field of the order of a few tesla. Studies on the optimal alloy composition pointed to Ti-rich composition for enhanced pinning, similar to the alloys normally employed for the fabrication of cables for superconducting magnets, Nb-47wt\%Ti (47\% titanium by weight) --- equivalent to Nb$_{0.37}$Ti$_{0.63}$ in atomic composition~\cite{Larbalestier:1989}.

The RADES collaboration has already experience with superconducting coatings~\cite{Golm:2021ooj}, either of Nb$_{3}$Sn or HTS, that led to the  first axion physics papers making use of HTS-enhanced cavities~\cite{Ahyoune:2024klt}. The reported Nb$_{3}$Sn coating was carried out at the CERN thin film coatings facilities. A preliminary test of the capabilities of realizing an NbTi coating has also recently been carried on the same type of $\sim$9~GHz cavity used for the previous RADES experiments~\cite{Golm:2021ooj}. For practical reasons, the coating has been realized with a stock source target of a Nb-rich composition (Nb$_{0.53}$Ti$_{0.47}$ in atomic ratio). This alloy was optimized for SRF accelerating cavities, and it is expected to provide a lower pinning than Ti-rich alloys, which are optimized for operation in a strong magnetic field.  First results reported at~\cite{DQLF_NbTi_coating} showed a quality factor at 3~T and $\sim$9~GHz equal to the value of copper. The Nb-rich alloy thus does not allow reaching the $\times 10$ increase reported for the QUAX study, however even with this sub-optimal composition it is expected that at frequencies around 350~MHz the gain should be of the order of $\times 100$, given the different dependencies of the surface resistance of copper ($\omega ^{1/2}$) and of NbTi ($\omega ^{2}$) under a magnetic field~\cite{Vaglio:2017}, for frequencies lower than the pinning frequency, which is the case in our conditions~\cite{Alimenti2022}. Current plans foresee testing also a Ti-rich alloy coating using an adapted Ti-rich target that has been recently procured, on the same type of 9~GHz cavity. The main objective of the study is to test the quality of the coating obtained with a sputtering process performed at low temperature. Performing coatings at high temperatures (500$\div$550~\degree C), such as those demonstrated in~\cite{Braggio2024}, has the advantage of promoting the formation of the correct $\beta$-NbTi phase together with $\alpha$-Ti precipitates that enhance magnetic field pinning~\cite{alpha-NbTi}. However, performing a high-temperature coating on objects of the size of the DQLF cavities presents formidable technological challenges for the design and realization of the UHV coating envelope. It should in principle be possible to perform a room-temperature coating (temperature increases naturally to 100$\div$150~\degree C due to the coating power) in one of the coating systems existing in the CERN coating laboratory without any major modifications, requiring only the need of purchasing a sputtering source of adequate size, and sufficient manpower availability for the coating development. We finally note that, similarly to what was done for previous RADES and QUAX coatings, the extremities of the cavity might be left without a superconducting coating,  depositing NbTi only onto the ``cylindrical" surface, greatly alleviating the difficulties that may arise in obtaining a correct composition and film structure on surfaces which are geometrically very different, while still allowing a significant improvement in the quality factor, according to the results obtained in the simulation.


\subsection{SQUID-Based Low-Noise Readout}
\label{sec:squid_readout}

The sensitivity of a microwave haloscope is fundamentally determined by its signal-to-noise ratio (SNR) as given by the Dicke radiometer equation,
\begin{equation}
    \text{SNR} = \frac{P_{\text{sig}}}{k_B T_{\text{sys}}} \sqrt{\frac{t}{\Delta f}},
    \label{eq:dicke_radiometer}
\end{equation}
where \(P_{\text{sig}}\) is the expected axion-induced signal power, \(T_{\text{sys}}\) is the system noise temperature, $\Delta f$ is the signal bandwidth, \(t\) is the integration time, and \(k_B\) Boltzmann's constant.	Consequently, the scan rate scales approximately as \(T_{\text{sys}}^{-2}\)	making the reduction of receiver noise one of the crucial requirements for any competitive axion search experiment.
Improvements in system noise temperature directly translate into drastically faster scanning of parameter space and improved sensitivity.

For this reason, the receiver chain within the DQLF programme focuses on the implementation of a superconducting quantum interference device (SQUID) based first-stage amplifier.
Whereas state-of-the-art cryogenic semiconductor amplifiers are typically limited to noise temperatures of a few kelvins at sub-GHz frequencies, the microstrip SQUID amplifiers (MSAs), with the exceptional flux sensitivity of the SQUIDs exploiting the Josephson effect and flux-quantization, can operate near the quantum limit.
Noise temperatures as low as \(\sim 50\,\text{mK}\) have been demonstrated at 519 MHz \cite{RevSciInstrum.92.124502, ApplPhysLett.78.967}, i.e.\ close to the quantum limit, which is \(\sim 25\,\text{mK}\) at 519 MHz, hence substantially outperforming conventional transistor-based technologies.
For the DQLF cavities, designed to cover a combined frequency range of approximately \(290\text{--}450\,\text{MHz}\), it is in principle possible to benefit from the correspondingly lower quantum-limited noise temperature of about \(14\text{--}22\,\text{mK}\). Operation below \(400\,\text{MHz}\), however, introduces additional experimental challenges, most notably the need to maintain sufficient RF isolation between successive amplification stages and to suppress the noise contribution of the subsequent amplifier.

For MSAs designed for operation around 519~MHz~\cite{ApplPhysLett.78.967}, gains of approximately 20~dB have been demonstrated together with input-referred noise temperatures below 100~mK. At this level of MSA gain, the noise contribution from the subsequent amplification chain is expected to remain below, or at least comparable to that of the MSA itself, as conventional (cryogenic) amplifiers based on high-electron-mobility transistor (HEMT) have sufficiently low noise.
To assess these aspects for the present experiment and characterize the performance of MSAs specifically optimized for operation around \(300\,\text{MHz}\), we performed an initial characterization of their operating conditions, gain, and noise temperature. The latter was determined using a heated-load measurement, with the MSA operated at approximately \(10\,\text{mK}\) as the first amplification stage of a receiver chain comprising a cryogenic HEMT amplifier at 4~K followed by room-temperature amplifiers, as described below.

\subsubsection{Microstrip SQUID Amplifier design}
\label{sec:msa_design}
For the initial measurements with the prototype cavities, we employ MSAs developed by \textit{ez SQUID Mess- und Analysegeräte (M. Mück)} \cite{mueckWebpage}, optimized for operation close to 300 MHz. Future efforts will focus on developing more field-tolerant and broadband SQUID amplifiers, investigating GrAl, NbN or other material options, tailored to this low-frequency regime.

The MSAs combine the high flux sensitivity of a shunted DC SQUID with a resonant microstrip input circuit.
Its noise performance is mainly determined by the self-flux parameter \(\beta_L\), which governs magnetic-flux screening and allows maximum flux-to-voltage transfer and sensitivity for \(\beta_L \sim O(1)\), and the hysteresis parameter \(\beta_c\), which controls the junctions’ non-hysteretic behaviour allowing convenient readout without bi-stability (or random telegraphic noise) for \(\beta_c < 1\).
The non-hysteretic DC SQUIDs developed by M. Mück and co-workers use Nb-Al\(_x\)O\(_y\)-Nb tunnel junctions with critical currents of a few \SI{}{\uA} \cite{ApplPhysLett.78.967}.
In the conventional Nb washer design with an 11-turn Nb input coil, the obtained parameters were \(\beta_L \approx 2.5\) and \(\beta_c \approx 0.8\).
This yields a high flux-to-voltage transfer coefficient $V_{\Phi} = |dV/d\Phi| \sim$ \SI{100}{\uV}$/\Phi_0$, where \(\Phi_0 = h/2e\) is the flux quantum, and flux noise as small as $S_{\Phi} \sim$ \SI{3}{\micro\relax}$\Phi_0\text{Hz}^{-1/2}$ \cite{ApplPhysLett.78.967}.

The operation bandwidth of the conventional DC SQUID (without the microstrip design) is primarily limited by the parasitic capacitance between the Nb square washer of the SQUID loop and the input coil, which behave like a parallel plate capacitor to ground with very limited tunable range, essentially making the device behave like a low pass (\(\sim\)DC) amplifier.
To use these conventional SQUIDs as a sub-GHz frequency amplifier, the microstrip SQUID architecture is implemented, where the input coil and SQUID washer form a resonant microstrip structure.
In this architecture, the limitations due to the parasitic capacitance were overcome by connecting the microwave signal source between the innermost turn of the input coil and ground while keeping the counter electrode at the opposite side (to the Nb square washer) of the tunnel junctions grounded (Fig.~\ref{fig:squid_noise_temp}a).
In this way, the parasitic capacitance between the Nb square washer is turned into a series capacitor with the capacitance largely tunable according to the length of the input coil.
Modifying the length of the input coil, hence the SQUID can be used as a bandpass amplifier with tunable band or working frequency range.
Depending on the length and number of turns of the input coil optimized for \(\sim 300\,\text{MHz}\) operation, the \(\beta_L\) and \(V_{\Phi}\) differs from the above-mentioned values, but eventually the flux noise \(S_{\Phi}\) is expected to exhibit similar order or magnitude.

\subsubsection{MSA operation}
\label{sec:msa_operation}
\begin{figure}[htb]
    \centering
    \includegraphics[width=0.98\linewidth]{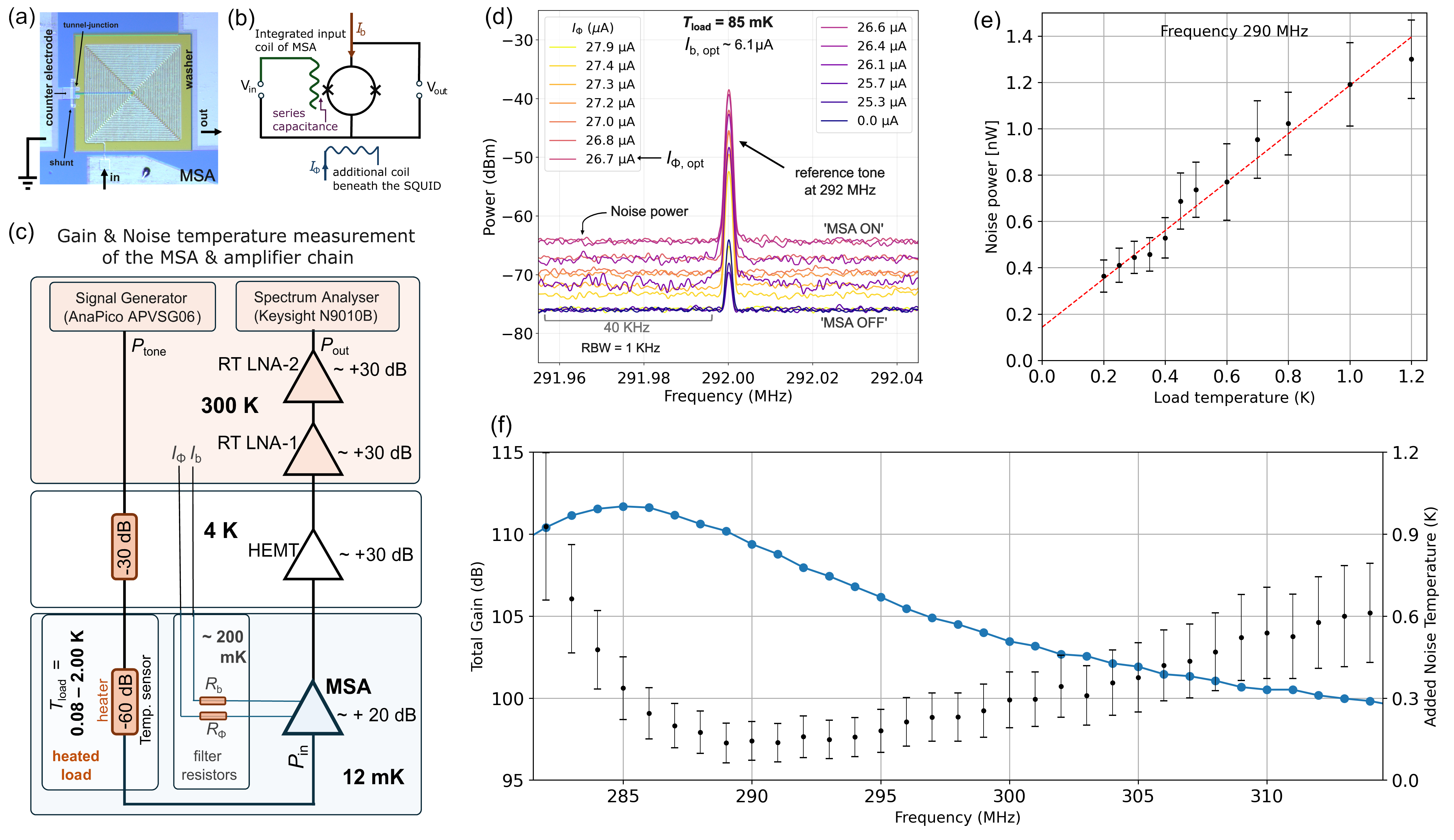}
    \caption{(a) Photograph and (b) schematic of the 300 MHz microstrip SQUID amplifier (MSA). (c) Cryogenic setup used for gain and noise-temperature measurements employing a variable-temperature Johnson-noise source. (d) Optimization of the SQUID flux-bias current using a weak calibration tone at 292 MHz, showing maximum gain near the optimum operating point. (e) Determination of the total receiver gain and added noise temperature from the measured output noise power as a function of load temperature at 290 MHz. (f) Frequency dependence of the total receiver gain and added noise temperature, yielding a minimum added noise temperature of approximately 0.1--0.2 K near 290 MHz. Measurements were performed without a cryogenic isolator between the MSA and HEMT amplifier.}
    \label{fig:squid_noise_temp}
\end{figure}

As described above, the counter electrode on the side opposite the Nb washer is held at ground potential, whereas the washer is biased through a filter resistor, which is weakly thermalized with respect to the SQUID (Fig.~\ref{fig:squid_noise_temp}c).
Typically, a DC bias current \(I_{b}\) \SI{<10}{\uA} is sufficient to drive the SQUID into its non-hysteretic, flux-sensitive voltage regime.
While the microwave input signal is coupled through the integrated input coil (Fig.~\ref{fig:squid_noise_temp}a), an additional on-chip coil located beneath the SQUID loop provides an independently controllable magnetic flux bias (Fig.~\ref{fig:squid_noise_temp}b).
Passing a current \(I_{\Phi}\) through this coil generates an offset flux that allows the SQUID to be biased at the steepest slope of its periodic characteristic, where the flux-to-voltage transfer function is maximized.

The operating point of the microstrip SQUID amplifier is therefore defined by two control parameters: the SQUID bias current \(I_{\text{bias}}\) and the flux-bias current \(I_{\Phi}\).
In practice, \(I_{\text{bias}}\) is adjusted slightly above the critical current of the Josephson junctions, while \(I_{\Phi}\) is tuned to position the SQUID at its point of maximum flux sensitivity, i.e.\ steepest slope of its periodic characteristic.
Optimization is typically performed by injecting a weak calibration tone into the amplifier and monitoring the output on a spectrum analyzer.
The two bias parameters are then iteratively adjusted to maximize the microwave gain.
Because unavoidable asymmetries between the Josephson junctions introduce a weak coupling between current and flux bias, several successive adjustments of both parameters are generally required to reach the true optimum.
For example, one such set of spectra while optimizing \(I_{\Phi}\) is presented in Fig.~\ref{fig:squid_noise_temp}d for a fixed \(I_{\text{bias}}\) close to its optimum value.
The optimum corresponds to the spectrum with maximum gain in both the reference tone and noise floor, as pointed out in Fig.~\ref{fig:squid_noise_temp}d.

In the present implementation, the MSA is operated through this two-parameter optimization procedure without active feedback.
The optimum bias currents \(I_{\text{bias}}\) and \(I_{\Phi}\) were found to depend on the operating frequency, SQUID temperature, and magnetic-field history, including possible flux trapping.
Consequently, periodic re-optimization is required to maintain maximum gain and minimum noise temperature.
An automated feedback system for continuous bias stabilization is currently under development for integration into the full DQLF receiver chain.

\subsubsection{MSA gain and noise-temperature characterization}
\label{sec:msa_characterization}

Following the optimization procedure described above, the noise performance of the MSA and the complete receiver chain was characterized in a dedicated cryogenic test configuration.
The measurement setup is shown in Fig.~\ref{fig:squid_noise_temp}c.
The resonant cavity was replaced by a calibrated thermal noise source consisting of a cryogenic attenuator mounted on a weakly coupled temperature-controlled stage.
The attenuator acted as a tunable Johnson-noise source whose physical temperature \(T_{\text{load}}\) could be varied between 80 mK and 2 K while the remainder of the receiver chain remained close to its nominal operating temperature.

Particular care was taken to avoid heating of the MSA by the DC connections.
The bias-filter resistors \(R_{\text{bias}}\) and \(R_{\Phi}\), required for supplying the SQUID bias and flux currents, were therefore thermalized at an intermediate stage near 200 mK rather than at the mixing chamber.
With superconducting wires connected between the bias resistor and the SQUID, no measurable increase of the MSA bath temperature was observed during application of operating currents.

A weak reference/calibration tone at 292 MHz was injected together with the thermal noise source.
Representative spectra acquired with a resolution bandwidth of 1 kHz are shown in Fig.~\ref{fig:squid_noise_temp}d.
As the flux-bias current \(I_{\Phi}\) approaches its optimum value, both the reference-tone amplitude and the noise floor increase simultaneously, indicating maximum MSA gain.
Comparison of the `MSA ON' and `MSA OFF' spectra provide early indication that the noise floor is strongly contributed by the MSA as desired.

The receiver gain and added noise temperature were extracted using the Y-factor method \cite{Engen1970, Kutlu_2021} .
For a linear receiver, the measured output noise power \(P_{\text{out}}\) is \cite{Friis1944}:
\begin{equation}
    P_{\text{out}} = G_{\text{total}}  k_B  \Delta f  (T_{\text{load}} + T_{\text{add}}),
    \label{eq:y_factor_output}
\end{equation}
where \(G_{\text{total}}\) is the total gain of the amplification chain, \(T_{\text{load}}\) is approximately the variable-temperature of the load, and \(T_{\text{add}}\) is the equivalent receiver-added noise temperature projected at the input of the primary amplifier (MSA).
Consequently, the slope of \(P_{\text{out}}\) versus \(T_{\text{load}}\) yields the total gain \(G_{\text{total}}\), while the intercept provides \(T_{\text{add}}\).
The resulting linear dependence measured at 290 MHz is shown in Fig.~\ref{fig:squid_noise_temp}e.

The total added noise temperature follows the Friis relation \cite{Friis1944},

\begin{equation}
    T_{\mathrm{add}}
    =
    T_{\mathrm{SQ}}
    +
    \frac{T_{\mathrm{HEMT}}}{G_{\mathrm{SQ}}}
    +
    \frac{T_{\mathrm{RT1}}}
    {G_{\mathrm{SQ}}G_{\mathrm{HEMT}}}
    +
    \frac{T_{\mathrm{RT2}}}
    {G_{\mathrm{SQ}}G_{\mathrm{HEMT}}G_{\mathrm{RT1}}}
    +\cdots ,
    \label{eq:friis}
\end{equation}

where $T_{\mathrm{SQ}}$ denotes the intrinsic noise temperature of the SQUID amplifier and the remaining terms correspond to the subsequent amplification stages. The importance of a high-gain SQUID preamplifier is immediately evident. For a measured SQUID gain of $G_{\mathrm{SQ}}\approx 20~\mathrm{dB}$ ($\approx 100$ linear), a typical HEMT noise temperature of $T_{\mathrm{HEMT}}\approx 1.5~\mathrm{K}$ contributes only $T_{\mathrm{HEMT}}/G_{\mathrm{SQ}} \approx 15~\mathrm{mK}$, while a room-temperature amplifier with $T_{\mathrm{RT1}}\approx 200~\mathrm{K}$ and $G_{\mathrm{HEMT}}\approx 30~\mathrm{dB}$ ($\approx 1000$ linear) contributes only $T_{\mathrm{RT1}}/(G_{\mathrm{SQ}}G_{\mathrm{HEMT}})\approx 2~\mathrm{mK}$. The noise contributions of all downstream amplifiers are therefore strongly suppressed, such that the measured receiver noise approaches the intrinsic MSA noise temperature.

A linear fit to the data in Fig.~\ref{fig:squid_noise_temp}e yields a total gain of \(G_{total} = 110\,\text{dB}\) and an added noise temperature of \(T_{\text{add}} \approx 120\,\text{mK}\) at 290 MHz. 
Repeating the above analysis for measurements centred at different frequencies yields the frequency dependence of both $G_{total}$ and $T_{\mathrm{add}}$, as summarized in Fig.~\ref{fig:squid_noise_temp}f.
We find that across the 283–313 MHz band, the receiver gain remains above 100 dB, whereas the minimum added noise temperature reaches approximately 0.1–0.2 K near 290 MHz.
Outside this region, and to a lesser extent within it, systematic degradations in both gain and noise temperature are observed.
These deviations are consistent with standing-wave resonances arising from impedance mismatches within the measurement chain, especially between the MSA and HEMT, which modify the effective gain seen by the receiver and can locally elevate the measured added noise temperature.

The measured system noise temperature demonstrates that SQUID amplification already enables operation deep in the sub-kelvin regime, validating the chosen readout concept for DQLF.
Importantly, these measurements were performed without a cryogenic isolator between the SQUID amplifier and downstream amplifier stages.
In this configuration, noise originating in downstream amplifier stages can partially propagate back toward the SQUID input, increasing the effective system noise temperature.
As demonstrated in other haloscope receivers, proper cryogenic isolation suppresses reverse-propagating noise and provides impedance matching between first and second stage amplifiers.
We therefore expect that the addition of a dedicated cryogenic isolator will further reduce the effective system noise temperature, potentially approaching the 50 mK added noise level across the operating band.

The sensitivity of the MSA to bias parameters, together with the observed variations in gain and noise temperature in this reduced receiver configuration, implies that for extended axion-search campaigns the amplifier performance must be re-evaluated following significant perturbations such as thermal cycling, magnetic-field changes, or flux trapping, which can shift the optimum bias point.
Because the scan rate depends sensitively on the total system noise temperature, even a relatively modest increase in \(T_{\text{add}}\), for example from 50 mK to 100 mK, can require substantial re-optimization of the antenna-coupling parameter \(\beta\) to maintain comparable scan speed and sensitivity.
Continuous monitoring of the SQUID operating point, together with active bias stabilization and automated noise calibration, will therefore be important components of future DQLF operation and are planned for implementation in the final receiver configuration.

\subsection{Data acquisition chain}
\label{sec:daq}

The data recorded during a run fall into three categories. \emph{Physics} data carry the possible axion signal on top of the thermal background and set the raw sensitivity of the search. \emph{Calibration} data are acquired by injecting known signals into the receiver and analysing its response; they calibrate the
chain and quantify its sensitivity to axion dark matter. \emph{Auxiliary} data (cryogenic temperatures, bias points, positioner states, environmental sensors) do not enter the sensitivity directly but verify correct operation and flag anomalies.
Physics and calibration data share most of the radio-frequency (RF) chain and differ only in which source drives the chain and where the output is routed. The acquisition system must therefore switch efficiently between a small number of well-defined operating modes. Auxiliary data originate from many independent sensors and are read by a separate slow control system. 

\begin{figure}
    \centering
    \includegraphics[width=0.65\linewidth]{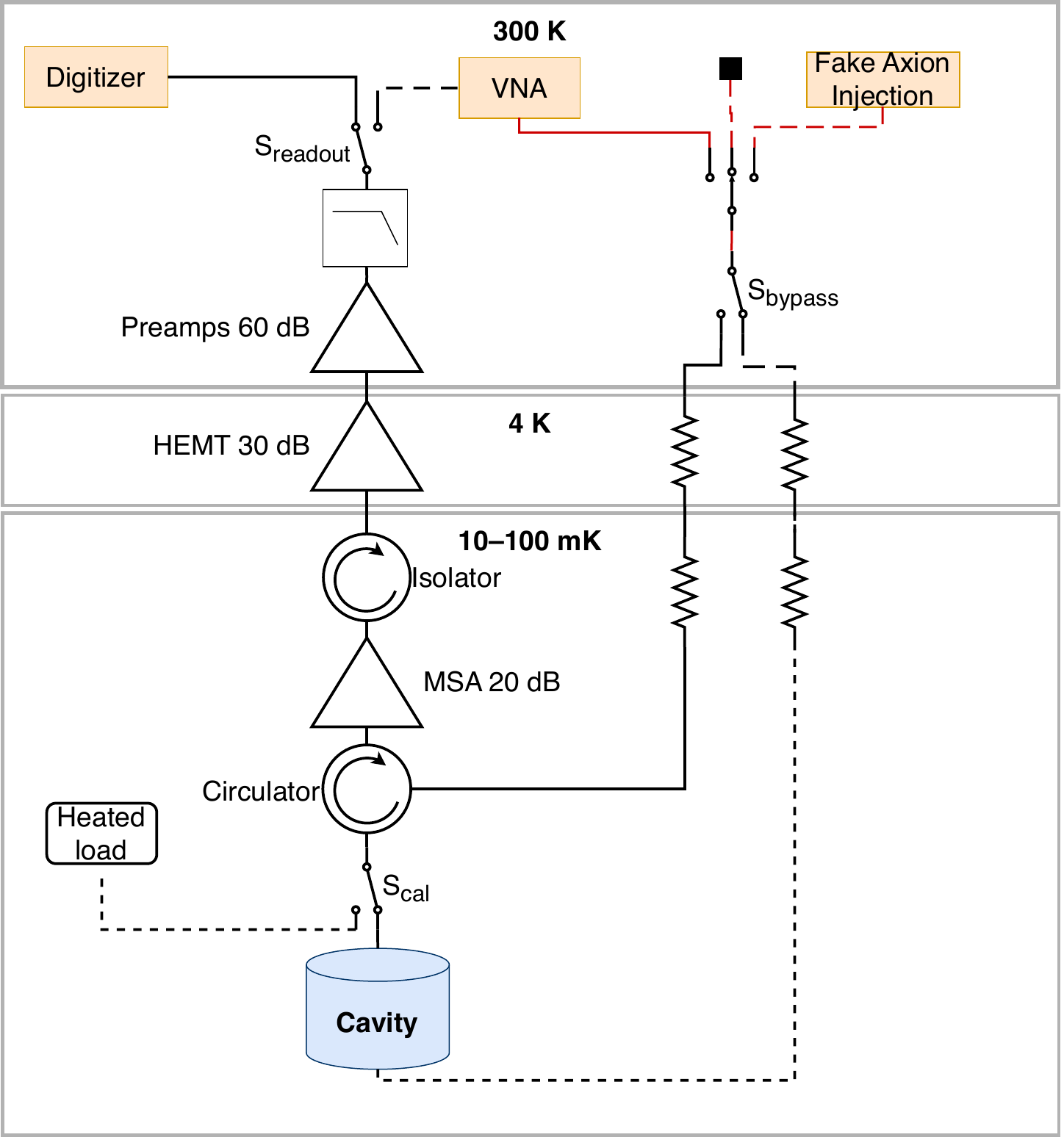}
    \caption{
Baseline architecture of the DQLF receiver and data-acquisition chain. The cavity signal is amplified by a microstrip SQUID amplifier (MSA) at the millikelvin stage, followed by a cryogenic HEMT amplifier and room-temperature amplification, before digitisation and spectral processing with the RFSoC digitizer. The cryogenic switching network provides alternative signal paths for noise calibration with a temperature-controlled heated load, characterization of the amplification stages, and fake axion injection. A VNA connection enables measurement of the cavity and receiver RF response.
}
    \label{fig:rf_chain_schematic}
\end{figure}

An overview of the RF chain is shown in Fig.~\ref{fig:rf_chain_schematic}. The
axion signal and the thermal background both originate in the tunable cavity, held at the mixing-chamber stage ($10$--$100\,\mathrm{mK}$). They are amplified first by the MSA, then by a cryogenic HEMT amplifier at the $4\,\mathrm{K}$ stage, and finally by two room-temperature amplifiers before digitisation. Physics data are sampled and processed by a radio-frequency system-on-chip based (RFSoC) platform. The RFSoC includes a digital RF Front which contains, among others, amplifiers, filtering stages, fast analogue-to-digital (ADCs) and digital-to-analogue (DACs) converters, programmable logic to implement accelerator units and several processors for real-time processing.

Calibration data are taken by injecting signals into the RF chain and measuring the response. In the current setup, a vector network analyser (VNA) drives the cavity through either the weakly coupled (transmission) or strongly coupled (reflection) port and reads the returning signal back through the amplifier chain, yielding the resonance frequency, the loaded quality factor $Q_{\rm l}$ and the antenna coupling $\beta$. A separate synthesiser can inject a synthetic (``fake-axion'') narrow-band tone into the cavity, which is then recovered through the full chain to validate end-to-end sensitivity. Mode selection is controlled by a set of RF switches that reconfigure the signal routing through the slow control.

\subsubsection{Digitisation and data taking}
\label{sec:data_taking}

The amplified signal and noise are sampled by the RFSoC ADC. The data is processed and digitally downconverted to baseband resulting in complex-valued in-phase/quadrature (I/Q) samples in fixed point arithmetic. To spectrally resolve a candidate signal, the record must be sampled continuously and coherently for longer than the axion coherence time, expected to be $2$--$5\,\mathrm{ms}$ in the DQLF frequency range. Coherent integration beyond the coherence time does not further improve the signal-to-noise ratio (SNR); additional sensitivity is gained instead by \emph{incoherent} integration as the noise present in the signal is supposed to be uncorrelated. In practice, each coherent record is Fourier-transformed into a power spectrum, and many such spectra are averaged: the noise power in each bin falls as $1/\sqrt{N}$ while the signal power stays constant, improving the SNR as $\sqrt{N}$ for $N$ averaged spectra.

Two acquisition strategies are being pursued:

\begin{itemize}
    \item {\textbf{Real-time spectral averaging.}}
Most haloscopes accumulate power spectra on the digitiser itself, since this compresses the data stream substantially. In the DQLF implementation the RFSoC programmable logic performs sampling, digital downconversion, decimation, the
FFT and magnitude-squared averaging on-chip, emitting one averaged power spectrum per readout together with a metadata record (see section~\ref{sec:data_analysis}). Existing firmware developed for higher-frequency haloscopes can be adapted to the DQLF band; this adaptation is in progress. The aim is to obtain a generic firmware able to serve any of the setups of the RADES collaboration, independently of its RF frequency that could ideally be parametrized in the firmware.
\item 
\textbf{{Time-domain streaming.}}
For DQLF the expected post-processed data rate is low enough that the time-domain I/Q stream can instead be written directly to disk, deferring all incoherent averaging to offline analysis. The data rate is set by the required span: a single power spectrum must cover at least a few times the cavity full width at half maximum (FWHM), which for $Q_L>10^{4}$ is at most $\Delta f \sim50\,\mathrm{kHz}$, requiring a sample rate of order of a few hundred kilohertz. Taking a representative sample rate of $500\,\mathrm{kSPS}$ and two bytes per quadrature yields a raw rate of $\approx 2\,\mathrm{MB\,s^{-1}}$ ($\approx 16\,\mathrm{Mbit\,s^{-1}}$), i.e.\ $\approx 0.17\,\mathrm{TB}$ per day.
Over one year of running at a duty cycle above $80\%$ this amounts to $\approx 50\,\mathrm{TB}$
of physics data. This volume is still readily handled with standard storage and represents a pessimistic upper limit. Any improvement in cavity Q-factor will proportionally reduce the data rate. Occasional broadband calibration and monitoring spectra (spanning $\sim100\times$ the cavity FWHM at reduced resolution) add only marginally to this total.

In the streaming mode, the only real-time processing required on the RFSoC is digital downconversion from the cavity resonance to baseband and decimation from the native ADC rate ($5\,\mathrm{GSPS}$) to a few $\times\,100\,\mathrm{kSPS}$; the comparatively expensive FFT is therefore no longer required in real time. Streaming also preserves considerable flexibility: because the raw I/Q record is retained, the spectral resolution can be chosen \emph{after} the run and adapted to different dark matter scenarios, including models predicting much narrower spectral features than the Standard Halo Model
~\cite{Yi:2023jrq,OHare:2023ajq}. In principle, the achievable frequency resolution is limited only by the duration of the recorded time series; in practice, the maximum coherent record length at a given tuning step is bounded by the dwell time $\Delta t$ at that cavity resonance. For the scan strategies considered here, even the shortest dwell times are at the $\sim100\,\mathrm{s}$ level (see section~\ref{sec:sensitivity}), corresponding to a Fourier frequency resolution of order $1/\Delta t\sim10^{-2}\,\mathrm{Hz}$ if the full record is analysed coherently. Shorter segments can of course be selected offline when a coarser resolution is appropriate.

\end{itemize}

\subsubsection{Calibration and monitoring}
\label{sec:calibration}

In order to relate the digitized power spectrum to the power output at the strong port of the cavity, the total gain of the receiver chain needs to be characterized. Equivalently, the system temperature referred to the input of the SQUID needs to be known. 

This is obtained by switching the SQUID input to a calibrated heated load; comparing the digitised power at different heated-load temperatures yields the calibration constant $C_\mathrm{cal}$ that converts raw ADC power to noise temperature, $T_\mathrm{sys}=C_\mathrm{cal}\,P_\mathrm{raw}$. The VNA measurements
provide the cavity parameters ($Q_\mathrm{l}$, $\beta$) which are required to relate the power output at the strong port of the cavity to a given axion coupling $g_{a\gamma}$. Fake-axion injection can provide an additional periodic end-to-end sensitivity check. System parameters ($T_\mathrm{sys}$, $Q_\mathrm{l}$, $\beta$, magnetic field, cryogenic temperatures, positioner states, run progress and alarms) will be logged continuously for monitoring by the slow-control subsystem.

\subsection{Cryogenic system}
\label{sec:cryogenic_system}

The cryogenic system will provide the low-temperature environment required for both the
resonant cavity and the quantum-limited receiver. The baseline objective is to cool the cavity
below $100\,\mathrm{mK}$ (target $<30\,\mathrm{mK}$), thereby reducing its thermal noise
contribution and enabling operation of the microstrip SQUID amplifier close to its optimum
noise temperature. The complete cryogenic installation will comprise the dilution unit,
gas-handling system, three cryocoolers, vacuum vessel, radiation shields, thermal links,
cryogenic wiring, and the mechanical interfaces required to support the one-metre-long cavity.

The cavity will be thermally connected to the lowest-temperature stage through
high-conductivity links designed to provide efficient cooling while limiting the transmission
of mechanical vibrations. Particular attention will be paid to the thermalisation of the cavity
walls, tuning plates, rotational axes, coupling antenna and coaxial connections. The large
dimensions of the cavity and its internal tuning system make control of thermal gradients
especially important. Temperature sensors will therefore be distributed at representative
positions on the cavity, thermal links and cryogenic stages in order to verify thermal
equilibrium during cooldown and data taking. Successive radiation shields and
intermediate-temperature stages will reduce the radiative heat load on the millikelvin stage.

The first-stage microstrip SQUID amplifier will be installed at the millikelvin stage and
connected to the cavity by a short, low-loss cryogenic coaxial line. The HEMT amplifier will
be mounted at the $4\,\mathrm{K}$ stage, while the remaining amplification and digitisation
electronics will be located at room temperature. Cryogenic RF switches and a calibrated
heated load will be incorporated to permit receiver calibration without warming the system.
All RF lines, SQUID-bias lines, tuning-system connections and thermometer wiring will be
carefully thermalised at the available temperature stages. Filtering, shielding and grounding
will be implemented to suppress electromagnetic interference and prevent additional heating
of the cold detector.

Because the SQUID performance is sensitive to magnetic-field history and flux trapping, the
amplifier will be installed in a lower-field region and protected by a dedicated magnetic-shielding
assembly. The final shield configuration will be validated experimentally at Karlsruhe Institute of Technology (KIT) before
installation at CERN. The cryogenic system will also include continuous monitoring of
temperatures, vacuum, cryocooler operation, gas pressures and cooling-water flow. Appropriate
alarms and interlocks will place the system in a safe state in the event of loss of cooling,
vacuum degradation or abnormal operating conditions.

The complete detector package will first be assembled and commissioned at KIT. This phase
will include cooldown tests, thermal-load measurements, operation of the tuning and coupling
mechanisms at low temperature, SQUID optimisation, receiver calibration and an extended
magnetless engineering run. The validated system will subsequently be prepared for transport
to CERN and recommissioned at the M1 facility. 

\section{Integration at CERN M1 magnet}
\label{sec:integration}

\subsection{M1 magnet}

\begin{figure}[htb]
    \centering
    \includegraphics[width=0.75\linewidth]{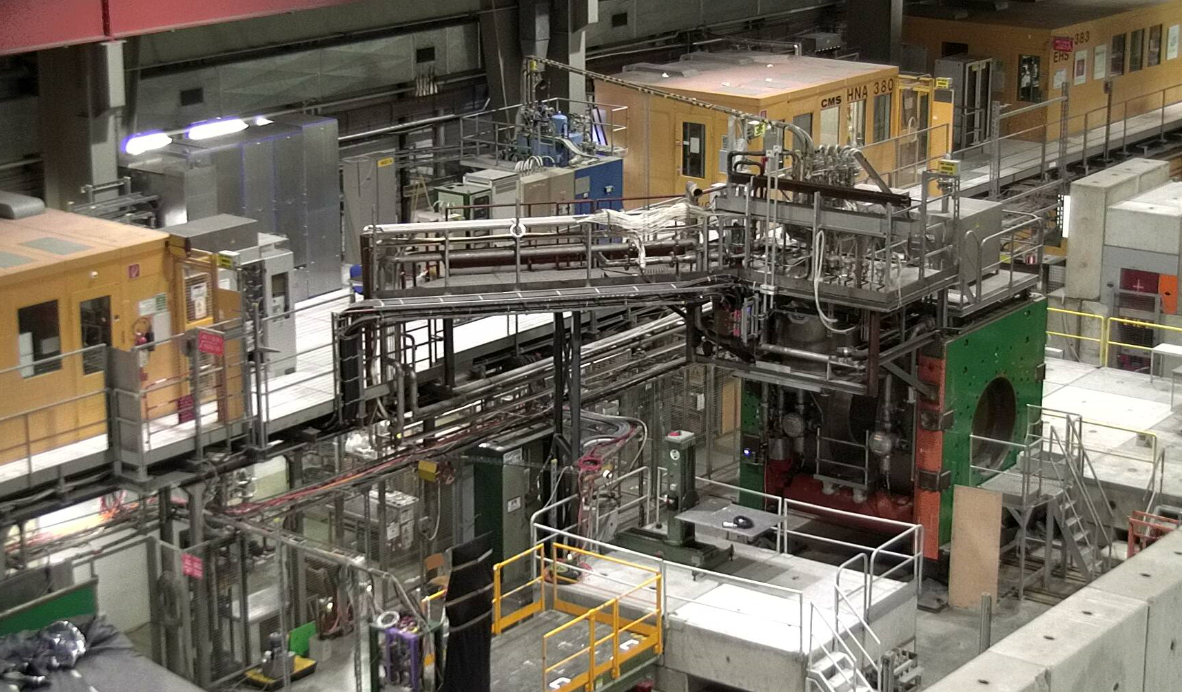}
    \caption{Birds-eye view of the M1 magnet and its cryogenic facility inside the EHN1 experimental area.}
    \label{fig:M1}
\end{figure}

The M1 magnet is a large-scale superconducting dipole facility located in the H2 test beam area within CERN’s North Area Experimental Hall (EHN1)~\cite{M1_magnet} (see Fig.~\ref{fig:M1}). Originally commissioned in 1980 as a central component of the European Hybrid Spectrometer (EHS) project, it initially supported fixed-target experiments such as NA16, NA22, NA23, and NA26. Since 1991, its primary mission has evolved into a vital calibration and testing resource for the Compact Muon Solenoid (CMS) experiment, allowing for the characterization of detector subsystems and auxiliary equipment in a high-field environment.

The magnet utilizes a Helmholtz coil configuration consisting of two NbTi superconducting windings. Each coil is housed in an independent cryostat and cooled to 4.2~K via immersion in a liquid helium bath. These cryostats are mounted on vertical steel plates and held together by four cylindrical steel rods designed to withstand the significant magnetic forces generated during high-field operation. The magnet has a total stored energy of 55~MJ.
M1 provides a nominal horizontal magnetic field (i.e. along the axis of the coils) of 3~T at its center, with the field strength reaching 3.6~T at the center of each coil. 

For standard operations at 3~T, the magnet requires a current of approximately 3980 A, and a full ramp-up to this level typically takes about two hours to ensure stability.
A primary advantage of the M1 facility is its exceptionally large central bore, which is maintained at room temperature. This volume is easily accessible from the top, the side, or along the coil axis, facilitating the integration of large-scale experimental setups. The key internal dimensions are:
\begin{itemize}
    \item Free internal diameter (bore): 1400~mm
    \item Free distance between cryostats: 820~mm
\end{itemize}
M1 is currently the only test magnet at CERN capable of providing a 3~T magnetic field within such a large bore. This unique combination of high field and large volume is essential for hosting the large-scale resonant cavities required for low-frequency axion searches.

\subsection{Experimental layout}
\label{sec:layout_fringe_fields}

\subsubsection{M1 magnet field maps}
\label{sec:M1-field-map}

\begin{figure}[h!]
    \centering
    \includegraphics[width=0.45\linewidth]{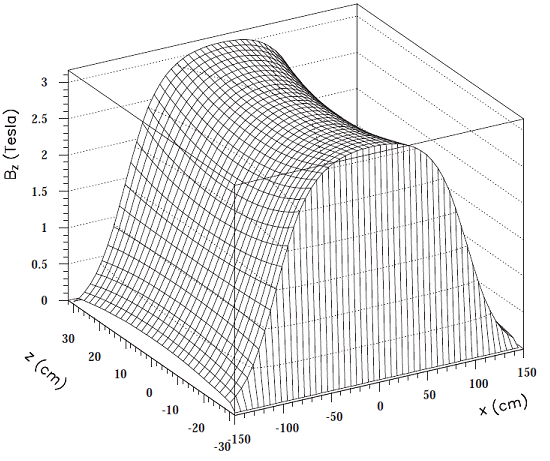}  \includegraphics[width=0.53\linewidth]{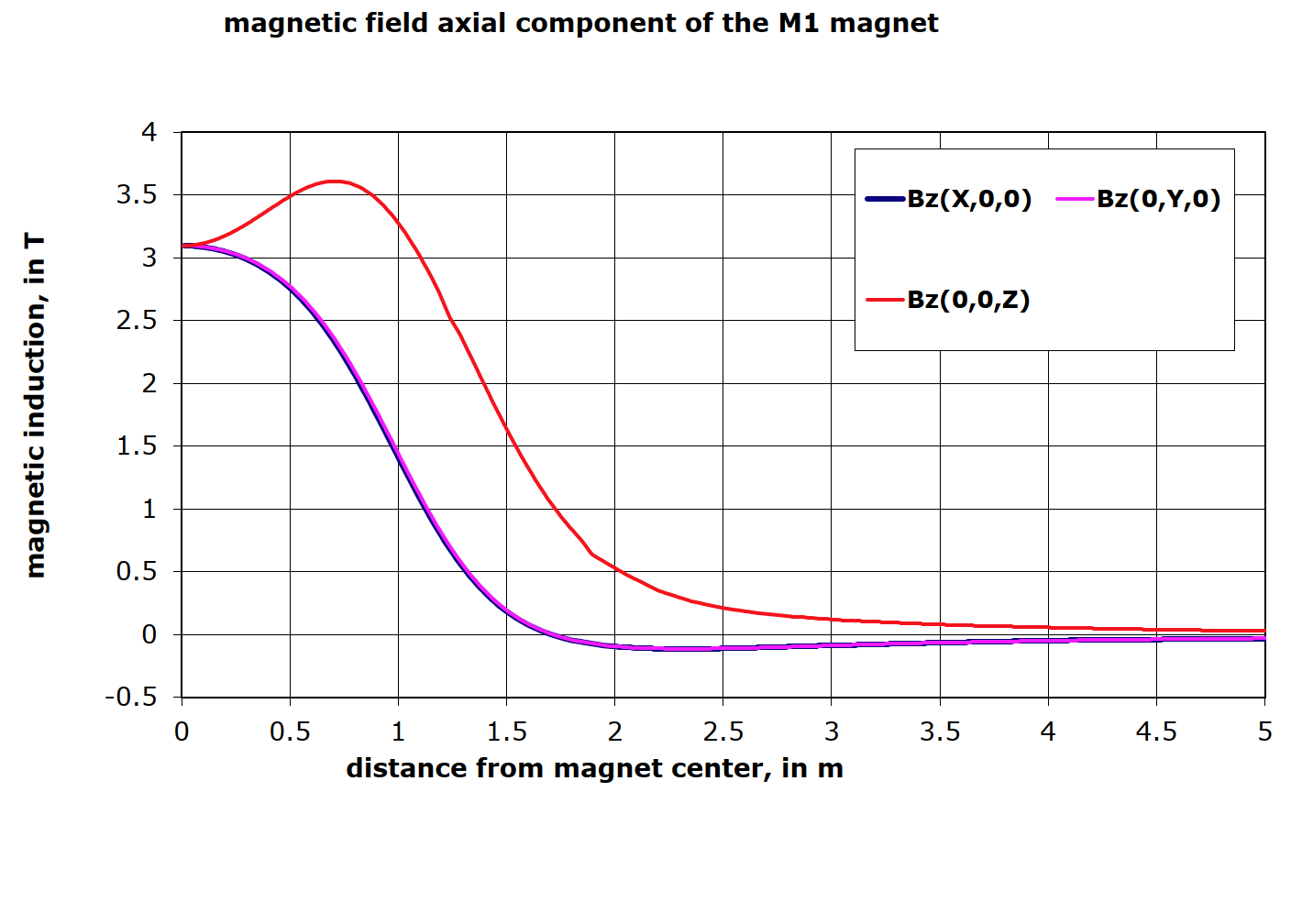}
    \caption{3D visualization of the axial field (\textit{left}) and plot of the axial field along the three main axes from the centre of the magnet (\textit{right})~\cite{M1_magnet}. The \textit{z} coordinate runs along the axis of the coils.}
    \label{fig:M1_field}
\end{figure}

The nominal field maps of the M1 magnet are discussed in detail in~\cite{M1_magnet}; we report in Fig.~\ref{fig:M1_field} a 3D visualization of the axial field and a plot of the axial field along the three main axes from the centre of the magnet, to support further discussions in this section.

\subsubsection{Optimal position within M1 magnet}
\label{sec:optimal-position}

The sensitivity of the proposed experiment also depends on the position of the cavity within the M1 magnet. Since the axion-induced signal power depends on the overlap between the resonant electric field and the external magnetic field, the spatial variation of the M1 field must be taken into account when determining the optimal cavity placement. This dependence is encoded in the form factor $C$ (see eq. \ref{eq:C}).

The relevant magnetic quantity entering the expected axion signal power is $C B_{\mathrm{avg}}^2$, where $B_{\mathrm{avg}}$ represents the effective magnetic field over the cavity volume. The signal power therefore scales as $P_a \propto C B_{\mathrm{avg}}^2$. Correspondingly, the scan rate scales as
\begin{equation}
    \frac{dm_a}{dt} \propto \left(C B_{\mathrm{avg}}^2\right)^2,
\end{equation}
while the reachable axion-photon coupling scales as
\begin{equation}
    g_{a\gamma} \propto \left(C B_{\mathrm{avg}}^2\right)^{-1/2}.
\end{equation}
Thus, spatial variations of the magnetic field and of its overlap with the cavity mode can directly affect both the scan rate and the ultimate sensitivity of the experiment.

The optimal cavity position was studied numerically by combining the electric-field distribution of the axion mode $\mathrm{qTE}_{111-\mathrm{y}}$, obtained from the electromagnetic cavity simulations, with the M1 magnetic-field map described above. In the region of interest, the magnetic field is predominantly oriented along the $z$-direction, and the axion conversion is therefore dominated by the overlap of the corresponding component of the cavity electric field with $B_z$.

\begin{figure}[htb]
    \centering
    \includegraphics[height=0.32\textwidth]{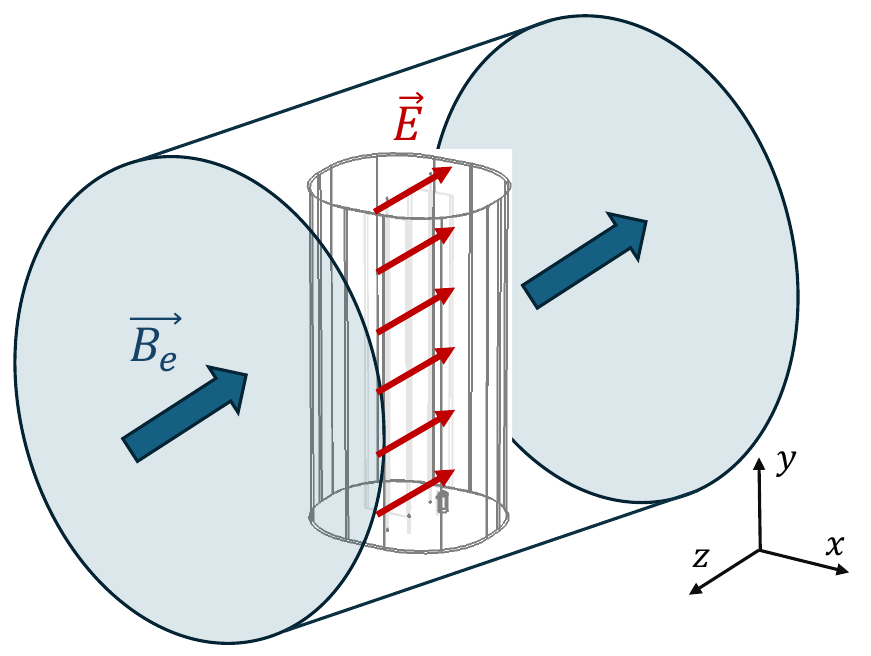} \hspace{5mm}
    \includegraphics[height=0.28\textwidth, trim={0cm 0 0cm 0},clip]{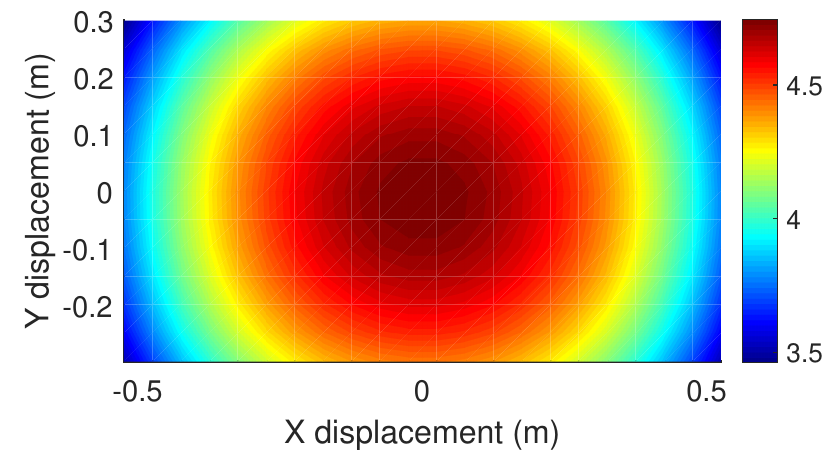}
    \caption{\textbf{Left:} Schematic view of the DQLF cavity inside the M1 magnet, with its longitudinal axis oriented along the $y$-direction. \textbf{Right:} Calculated magnetic figure of merit $C B_{\mathrm{avg}}^2$ as a function of the cavity displacement along the $x$- and $y$-directions at z = 0. The calculation combines the interpolated M1 magnetic-field map with the simulated electric-field distribution of the axion mode $\mathrm{qTE}_{111-\mathrm{y}}$, for a tuning-plate angle $\alpha=0^\circ$.}
    \label{fig:Form_factor_M1}
\end{figure}

The study considered transverse displacements of the cavity along the $x$- and $y$-directions, while keeping $z=0$. The cavity axis was oriented along the $y$-direction (and therefore, the electric field of the axion mode is oriented along the $z$-direction of the magnet reference system) and longitudinally centred in the high-field region. Figure~\ref{fig:Form_factor_M1} shows the geometry considered and the resulting value of $C B_{\mathrm{avg}}^2$ as a function of the cavity position. Over the range of displacements considered, the overlap remains relatively stable, with the variation in $C B_{\mathrm{avg}}^2$ being driven predominantly by the change in the effective magnetic field sampled by the cavity volume.

The maximum value of $C B_{\mathrm{avg}}^2$ is obtained when the cavity is centred at $(x,y,z)=(0,0,0)$. Nevertheless, the dependence on moderate transverse displacements is relatively weak. For example, displacements of $y=\pm0.3~\mathrm{m}$, with $x=z=0$, result in an approximate degradation of only $4\%$ in the reachable axion-photon coupling and $8\%$ in scan rate. The experiment therefore has some tolerance to transverse positioning along the $y$-direction, although the centred configuration remains clearly preferred.

The effect of the tuning system has also been considered for the centred cavity configuration. As the tuning plates rotate, the electromagnetic field distribution of the axion  mode evolves and consequently so does its overlap with the M1 magnetic field. The form factor must therefore be evaluated throughout the tuning range rather than represented by a single value. This dependence, discussed in Sec.~\ref{sec:cav_characterisation}, needs to be taken into account in the final sensitivity calculations.

The present analysis is based on an interpolation of the available M1 magnetic-field data. A dedicated field mapping of the magnet would therefore be valuable for the final sensitivity calculation, allowing a more accurate determination of the spatial overlap and of $C B_{\mathrm{avg}}^2$. Once available, the measured field map will be incorporated into the final calculation.

The cavity position must also be considered together with the magnetic-field requirements of the SQUID amplifier. While the cavity benefits from being located at the centre of the magnet, where $C B_{\mathrm{avg}}^2$ is maximised, the SQUID must operate in a sufficiently low residual magnetic field to avoid degradation of its performance. The preferred configuration is therefore to retain the cavity close to the magnet centre and locate the SQUID further away, connecting the two through an appropriate length of cryogenic coaxial cable. At the frequencies relevant for DQLF, the attenuation of suitable rigid coaxial lines is expected to be small, typically of order $0.01$--$0.1~\mathrm{dB/m}$ \cite{Kurpiers:2016vhq}. Cable lengths of approximately $0.5$--$2~\mathrm{m}$ therefore introduce only a modest contribution to the overall system noise compared with the sensitivity penalty that would result from displacing the cavity significantly from the optimal magnetic-field region.

The baseline configuration consequently places the cavity at, or as close as mechanically possible to, the centre of the M1 magnet, while positioning the SQUID in a lower-field region compatible with its magnetic requirements and with the mechanical and cryogenic constraints of the experiment. The detailed implementation of this configuration is discussed in the following sections.

\subsubsection{Fringe fields}
\label{sec:Fringe}

\begin{figure}[t]
    \centering
    \includegraphics[width=0.72\textwidth]{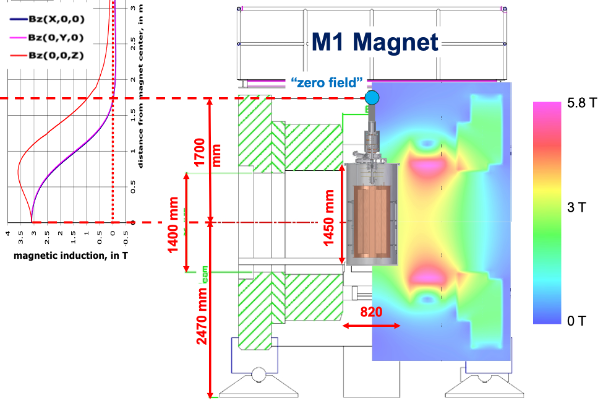}
    \caption{Schematic view of the M1 magnet and magnetic-field map. The cavity is located
    at the centre of the magnet. The ``zero-field'' region is indicated by the blue circle.
    Mu-metal and superconducting shields will additionally be used to minimise the magnetic
    field at the microstrip SQUID amplifier.}
    \label{fig:m1_field_map}
\end{figure}

The cryogenic detector will be positioned within the room-temperature bore of the M1 magnet,
with the resonant cavity located as close as mechanically possible to the centre of the magnetic
volume (see Fig.~\ref{fig:m1_field_map}). This position maximises the overlap between the
cavity electric field and the external magnetic field and therefore maximises the
axion-conversion power. 
The final alignment will be established from the CERN magnetic-field map and verified by an on-site geometrical survey.

The microstrip SQUID amplifier cannot be operated in the full central field and will therefore
be positioned outside the highest-field region, close to the ``zero-field'' region indicated in
Fig.~\ref{fig:m1_field_map}. It will be connected to the cavity by a low-loss rigid or
semi-rigid coaxial line. 
This permits the cavity to remain at the optimum magnetic position while the SQUID is placed in a region compatible with its shielding and operating requirements.

\begin{figure}[t]
    \centering
    \includegraphics[height=5cm]{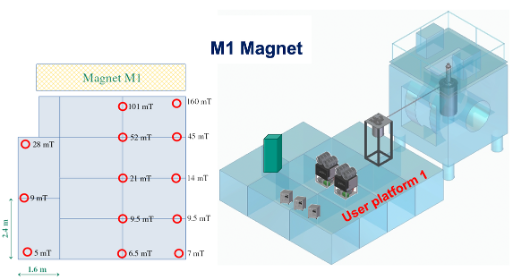}
    \includegraphics[height=4cm]{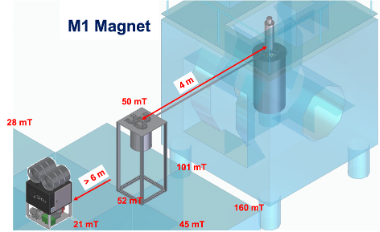}
    \caption{Fringe-field distribution of the M1 magnet and approximate positioning of the
    experimental equipment.}
    \label{fig:m1_fringe_fields}
\end{figure}

The complete fringe-field distribution will be used to define installation zones for the SQUID,
magnetic shielding, cryogenic equipment, tuning actuators, vacuum instrumentation and
room-temperature electronics. An approximate positioning of the experimental equipment is
shown in Fig.~\ref{fig:m1_fringe_fields}. Before final installation, the available field map
will be supplemented, where necessary, by direct field measurements at the intended
equipment positions. Components containing ferromagnetic materials, or devices whose
operation may be affected by magnetic fields, will be located outside their specified field
limits. Particular attention will be given to the SQUID magnetic shield, piezoelectric
actuators, RF electronics, temperature sensors and vacuum instrumentation.

The field-ramping procedure will be coordinated with the cryogenic and readout operation.
SQUID gain, bias point and noise performance will be checked after major magnetic-field
changes because flux trapping or changes in magnetic-field history can alter the optimum
operating conditions. 

\subsection{Mechanical integration}
\label{sec:mechanical_integration}

The mechanical integration will follow a modular approach in which the cryostat, cavity,
tuning mechanism, SQUID package and cryogenic services are assembled and qualified as a
complete detector unit at KIT before shipment to CERN. The support structure will position
the one-metre-long cavity reproducibly at the centre of the M1 magnetic volume while
remaining compatible with the $1400\,\mathrm{mm}$ free bore and the $820\,\mathrm{mm}$
space between the magnet cryostats. The design will also permit the three DQLF cavity
geometries to be exchanged between measurement periods without requiring modification of
the main cryogenic infrastructure.

\begin{figure}[t]
    \centering
    \includegraphics[width=0.75\textwidth]{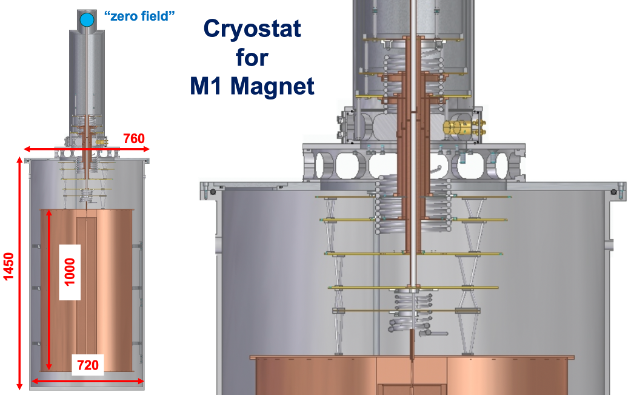}
    \caption{
Conceptual design of the DQLF cryostat and integrated one-metre cavity assembly. The resonant cavity is mechanically supported inside the cryostat and surrounded by the successive thermal stages required for operation at millikelvin temperatures.}
    \label{fig:dqlf_cryostat}
\end{figure}

The cryostat shown in Fig.~\ref{fig:dqlf_cryostat} will be a purpose-built, large-volume,
cryogen-free dilution-refrigerator system adapted to the dimensions and thermal mass of the
one-metre resonant cavity. To reduce development time and technical risk, the dilution unit
will make extensive use of established Qinu technologies\footnote{\href{Qinu: https://qinu.de/}{https://qinu.de/}} and engineering solutions,
particularly for the still, the continuous and step heat exchangers, the mixing chamber, and
the associated $^{3}$He-$^{4}$He circulation circuit. These elements will be integrated as a
coordinated cryogenic module rather than being developed independently from first principles,
facilitating reliable construction, leak testing, commissioning and subsequent maintenance.

The vacuum vessel, radiation shields, cavity suspension, thermal links and experimental
interfaces will nevertheless be designed specifically for the DQLF detector because of its
unusually large geometry and cold mass. The modular Qinu concept, in which the cryostat,
gas-handling system and cryocooler unit can be positioned separately, is particularly
advantageous for this experiment. Pumps and other vibration-producing components can be
located away from the cavity and SQUID amplifier and connected through flexible, thermally
insulated lines. This approach combines a custom cryostat tailored to the haloscope with
proven dilution-refrigeration technology, while preserving the flexibility required for
integration first at KIT and subsequently at the CERN M1 magnet.

The mechanical structure must support the static load of the cryostat and cold detector,
accommodate differential thermal contraction during cooldown, and maintain the alignment of
the cavity and tuning mechanism. It will be designed to minimise vibration transmission from
the cryocoolers and vacuum equipment to the resonator and SQUID. Materials used close to
the magnet will be non-magnetic and selected to minimise forces and heating induced during
magnetic-field ramping. Large closed conducting loops will be avoided where practicable, and
the cavity support, radiation shields and service connections will be checked for mechanical
stability under the expected magnetic and thermal conditions.

The tuning-plate shafts and coupling-antenna actuator will be integrated inside the cryostat at
$300\,\mathrm{K}$ in order to avoid vacuum-compatible rotary interfaces. Their supports will
maintain alignment over the full thermal cycle while providing the angular precision and
reproducibility required for the frequency scan. Cryogenic cables, coaxial lines, pumping
connections, cooling-water lines and electrical services will be routed with suitable strain
relief and with sufficient flexibility to accommodate cooldown and installation tolerances.
The arrangement will preserve access to the motors, vacuum components and readout interfaces
for maintenance.

A dedicated transport and lifting frame will be developed for movement of the integrated
detector from KIT to CERN and for installation using the EHN1 handling infrastructure.
Defined lifting points, mechanical stops and survey references will ensure safe handling and
reproducible positioning. Installation will proceed in stages: placement and alignment of the
support structure, insertion of the cryostat and cavity assembly, connection of cryogenic and
electrical services, vacuum and leak testing, cryogenic commissioning, magnet-off engineering
tests and, finally, controlled operation with the M1 magnet energised. The final mechanical
design, load calculations, lifting procedure and interface drawings will be reviewed together
with the relevant CERN engineering and safety teams before installation.

The proposal therefore foresees complete cryostat-cavity-SQUID integration and magnetless
qualification at KIT before shipment, followed by installation and cryogenic commissioning
at CERN.

\section{Data taking protocol and organization}
\label{sec:data-taking}

\begin{figure}[t]
    \centering
    \includegraphics[width=0.75\linewidth]{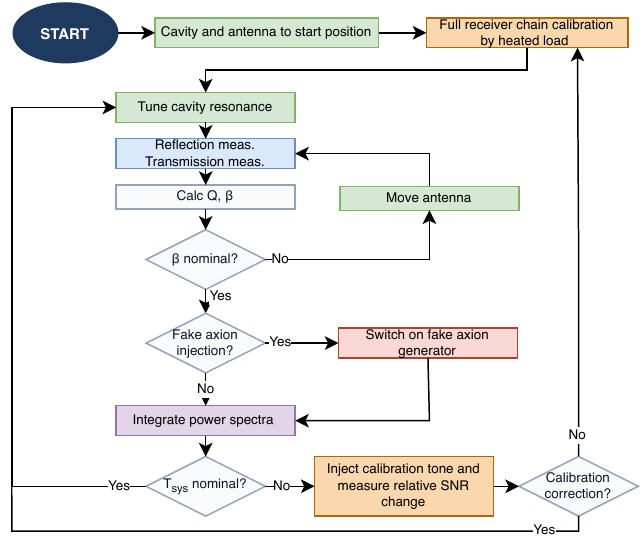}
  \caption{Flowchart of the proposed DQLF data-taking protocol. After initialization and a full calibration of the receiver chain using the heated load, each scan step consists of tuning the cavity resonance, measuring the reflection and transmission coefficients, and determining the cavity parameters $Q$ and $\beta$. The antenna position is adjusted when necessary to recover the nominal coupling. Physics data are then acquired for the prescribed dwell time, with optional fake-axion signal injection for validation. The stability of the receiver chain is monitored through periodic calibration-tone measurements, triggering a new full calibration when required. The sequence is repeated for successive cavity-frequency settings until completion of the run.}
    \label{fig:run_protocol}
\end{figure}

The planned data-taking sequence for DQLF is summarized in the flowchart of Fig.~\ref{fig:run_protocol}. The run will open with a setup phase, performed once: the tuning plates and the coupling antenna are driven to their start
positions, and the full receiver chain is calibrated
against a heated load to fix the absolute noise reference.
The experiment then enters the main acquisition loop shown in the sketch, which is repeated for every resonant-frequency step until the run is ended.

At the start of each iteration the cavity resonance is advanced by displacing
the tuning plates. Because the plates are moved mechanically, this step
comprises both the plate travel time and a subsequent dwell during which
residual vibrations damp out before any measurement is taken ($t_\mathrm{tune}$). Once the cavity is mechanically quiet, paired transmission ($S_{21}$) and reflection ($S_{11}$) sweeps are recorded in a characteristic time interval ($t_\mathrm{VNA}$), from which the loaded quality factor $Q_L$ and the antenna coupling $\beta$ are derived. If $\beta$ falls outside its nominal window, the loop antenna rotation is corrected; as with the plates, this involves an actuator movement followed by a
settling dwell ($t_\mathrm{ant}$), after which the $S$-parameter measurement is repeated. This correction branch may run for $n_\mathrm{ant}$ iterations before the coupling is accepted. As the cavity will have been characterized beforehand, it is deemed realistic that a single iteration could suffice in practice to recover an acceptable $\beta$.

Once the coupling is verified, a synthetic (``fake'') axion can be injected through the weakly coupled port to validate the entire receiver chain and analysis, possibly as a blind injection known only to a few individuals. The receiver output is then digitized and integrated into an on-resonance power spectrum over the acquisition time $t_\mathrm{int}$. This integration is the only interval in the cycle during which science data are
accumulated. The $\mathrm{SNR}$ improves as $\sqrt{N}$ where $N$ is the number of averaged spectra. For a given integration time $N= \eta_{\rm int} t_\mathrm{int} \Delta f$ where $\Delta f$ is the spectral resolution and $\eta_{\rm int}$ is the efficiency of the DAQ. Essentially zero-dead time ($\eta_{\rm int}\sim 1$) have been implemented before. However, folding in instrument communication delays into $t_{\rm int} $ will likely lead to $\eta_{\rm int}\sim 0.9$. 

Following each acquisition the system noise temperature
$T_\mathrm{sys}$ is assessed from the acquired data itself: if it remains nominal the protocol proceeds directly to the next frequency step, whereas a measured drift triggers the injection of a reference calibration tone and a measurement of the relative $\rm SNR$ change ($t_\mathrm{tone}$). Depending on that result, either an in-place correction is applied or the full heated-load calibration is repeated ($t_\mathrm{cal}$). The loop terminates once the run-ongoing condition is no longer satisfied. 

The efficiency of this protocol is set by its live-time duty cycle,
since every operation other than the integration $\eta_{\rm int}t_\mathrm{int}$ constitutes deadtime 
\begin{equation}
  t_\mathrm{dead} \simeq t_\mathrm{tune} + t_\mathrm{VNA}
    + \langle n_\mathrm{ant}\rangle\,\bigl(t_\mathrm{ant} + t_\mathrm{VNA}\bigr)
    + f_\mathrm{cal}\,t_\mathrm{tone}
    + p_\mathrm{recal}\,t_\mathrm{cal},
  \label{eq:deadtime}
\end{equation}
where $\langle n_\mathrm{ant}\rangle$ is the mean number of antenna corrections per step, $f_\mathrm{cal}$ the fraction of cycles requiring a tone check, and $p_\mathrm{recal}$ the probability of a full recalibration being triggered. From experience on the cavity prototypes, $t_{\rm tune}$ and $t_\mathrm{ant}$ are expected to be $\sim \SI{1}{\second}$ each. VNA measurements are expected to take a similar amount of time. The antenna coupling changes only slowly compared to the width of the resonance, therefore antenna retuning is expected occur rarely and $\langle n_\mathrm{ant}\rangle \ll 1$. Checking the calibration via a reference tone is expected to take on the order of a minute. A full heated load calibration takes on the order of an hour as this requires heating the load to several different temperatures and waiting for the system to equilibrate. In both cases there is still a large uncertainty on how often either will be required. Given an unavoidable dead time of already at least a few seconds from motor and VNA operations, one would require $f_\mathrm{cal}\lesssim 0.01$ and $p_\mathrm{recal}\lesssim 0.001$ to not contribute significantly to the total dead time.

The resulting overall duty-cycle efficiency is
\begin{equation}
  \eta = \frac{\eta_{\rm int}t_\mathrm{int}}
  {\eta_{\rm int}t_\mathrm{int} + t_\mathrm{dead}}.
  \label{eq:duty-cycle}
\end{equation}

Neglecting for the moment the additional dead time associated with periodic recalibrations, which will depend on the stability of the system during operation, and assuming $t_{\rm dead}=\SI{10}{\second}$ and $\eta_{\rm int}=0.9$, the overall duty-cycle efficiency remains above 0.82 over the range of scan conditions considered. The lowest efficiency corresponds to the limiting case with the largest loaded cavity quality factor, and therefore the narrowest resonance bandwidth and finest frequency spacing. This situation can arise, for example, for a high-$Q$ cavity operated with a circulator and without the additional bandwidth broadening associated with strong overcoupling. Since the total available data-taking time is fixed, the larger number of required tuning steps results in the shortest dwell time per step. In this limiting case, $\eta_{\rm int}t_\mathrm{int}=\SI{51}{\second}$, yielding $\eta\simeq0.82$. For the sensitivity scenarios discussed in Sec.~\ref{sec:sensitivity}, the lower loaded quality factor resulting from the adopted coupling conditions leads to longer dwell times and hence to a higher duty-cycle efficiency.

Candidate signals of interest will be revisited for further study after the run concludes. A more refined protocol, identifying candidates in real time and automatically cross-checking them to confirm or exclude an axion origin, is deferred to later runs, which should be more routine and easier to automate.
\section{Data treatment and analysis}
\label{sec:data_analysis}

The data treatment and analysis envisioned is based on a well established pipeline that follows mainly the analysis framework developed in the HAYSTAC collaboration \cite{Brubaker:2017rna}, which is considered to be the standard in the community for conventional haloscope searches. In the RADES collaboration we followed this analysis procedure broadly in previous data-taking campaigns \cite{CAST:2020rlf,Ahyoune:2024klt}, adapting it to handle the background noise of our acquisition system in a manner adapted to the electronics chain in use.

\begin{figure}[!h]
    \begin{center}
    \includegraphics[width=0.75\columnwidth]{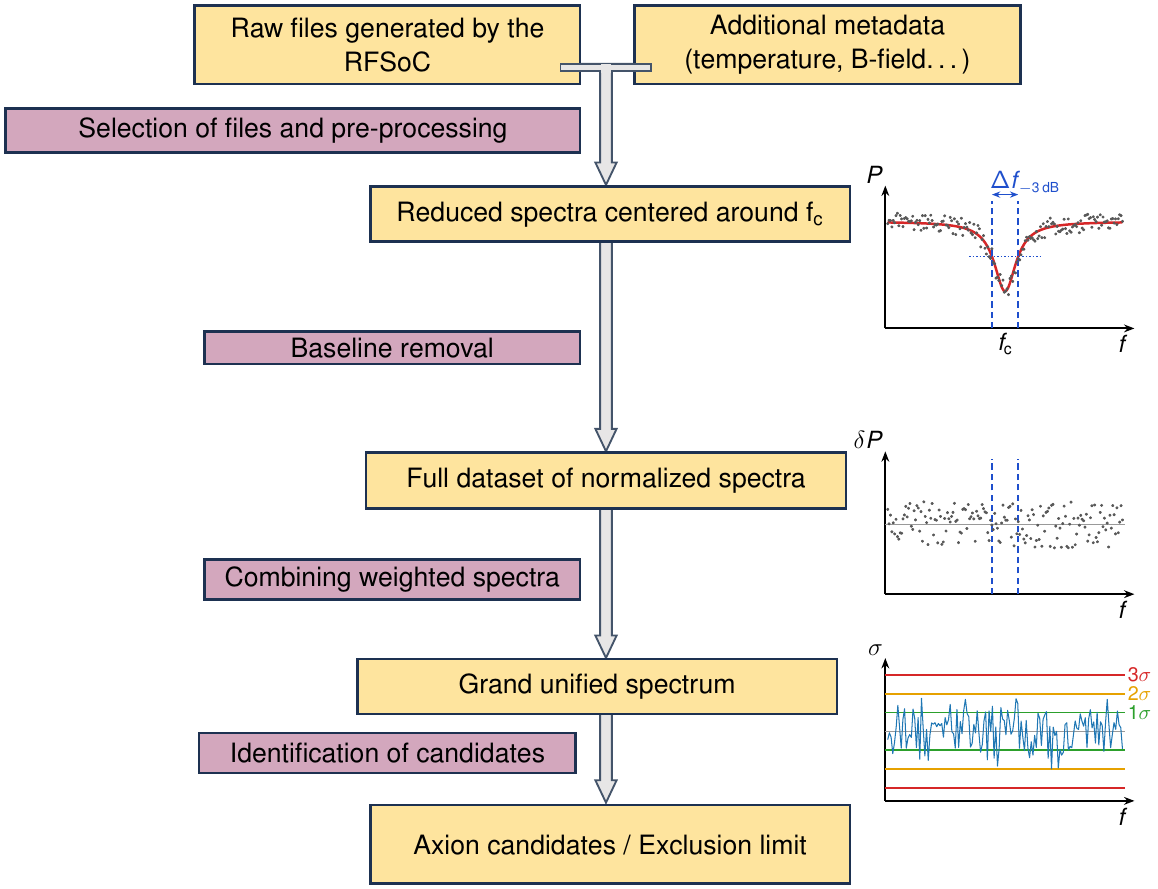}
	    \caption{Flowchart of the general structure of the analysis procedure proposed for the DQLF data. The main steps are similar to those established by the HAYSTAC collaboration \cite{Brubaker:2017rna} but the specific mathematical treatment of the data might differ slightly depending on the final data taking protocol. 
        } 
	    \label{fig:analysis_flow}
	\end{center}
\end{figure}

Figure~\ref{fig:analysis_flow} illustrates the general steps of the analysis pipeline followed in the most recent haloscope search performed by RADES, this analysis scheme can be directly applied to the DQLF setup. A, still general, but more detailed description of the different steps can be summarized as

\begin{enumerate}
\item \textbf{Pre-processing}: Data files are classified in subgroups according to metadata: physics run, calibration, background measurements, etc. Files with pathological behaviour, e.g. missing spectra, could also be tagged at this stage. If data reduction is needed, a reduced range of the spectra centred on the cavity resonance can also be generated in the same step.

\item \textbf{Baseline removal}: The behaviour of the baseline for each spectrum is characterised independently and subtracted, the resulting residuals are normalized with respect to that same baseline. This step should result on a series of spectra of Gaussian distributed noise. The usual approach to this step is the use of a Savitzky-Golay filter, however, other alternatives that include measurements of the electronics background itself could be studied.

\item \textbf{Combining spectra}: The normalized spectra are then combined in a weighted average where two factors are taken into account: the average noise power per bin and the profile of the cavity resonance for each independent spectrum. Depending on the final data taking protocol the profile of the resonance can be obtained either by independent VNA measurements or by fitting the spectra themselves.

\item \textbf{Identification of candidates}: According to a target statistical significance, a signal threshold is set. Then the grand unified spectrum is filtered by the axion lineshape in the frequency range covered by the experiment. This procedure would allow tagging candidate signals for a rescan to study their persistence. At the same time, the injected signals should be recovered in this step to verify the validity of the analysis procedure.

\item \textbf{Generating an exclusion limit}: If no persistent candidates are identified, based on the experimental parameters a exclusion limit in the frequency band covered by the experiment would be produced.

\end{enumerate}

It is worth noting that the first acquisition strategy described in \ref{sec:data_taking} would produce a time series of power spectra that can be processed exactly as described above without any additional steps. However, the second acquisition strategy requires a first preprocessing step. The raw time-series of I/Q values has to be Fourier transformed, and incoherently integrated to obtain the starting point of the previously described analysis pipeline. In this step the spectral resolution can be fixed to optimize sensitivity to the dark-matter model that one wants to test. From that point on, the analysis procedure would be the same.  Furthermore, recording the raw time-series itself would allow more sophisticated analysis of the same data: studying transients, correlation with other detectors, etc.

In order to facilitate a transparent, collaboration wide protocol for the analysis pipeline, the REST-for-Physics framework \cite{REST_TPCconf} has been expanded with the \texttt{halolib} library. The package has the possibility to initialize power spectra-type data objects alongside metadata dedicated to saving experimental parameters (such as the bandwidth, resolution, central frequency and integration time), select the physics region by trimming (as shown in panel (a) of Figure~\ref{fig:analysis_flow}), and combine spectra. 
As the REST framework is written utilizing ROOT particle physics software, the \texttt{halolib} library benefits from the extensive fitting procedures well documented in ROOT. This can be used early in the data analysis procedures to identify `bad runs' by a basic fit of the central peak (panel (b) of Figure~\ref{fig:analysis_flow}). 
Therefore, REST lends itself as a transparent, traceable coding environment for basic data analysis procedures used in the low level steps of the RADES analysis pipeline. Accessibility to this level of the data processing would also enable a new path for data sharing in both the collaboration and the wider axion community. 
\section{Sensitivity projections}
\label{sec:sensitivity}

The projected sensitivity of the proposed experiment has been estimated using the standard haloscope signal-to-noise formalism \cite{ADMX:2006kgb,Sikivie:1983ip}. The expected axion signal power is computed including the cavity-to-receiver coupling appropriate to each sensitivity scenario, as described below, while the exclusion limit at each frequency is obtained by requiring the signal corresponding to a chosen axion-photon coupling to reach the threshold signal-to-noise ratio corresponding to a 90\% confidence level.

The present calculation explicitly models a frequency scan over the full tuning range. The exclusion limit is evaluated independently at each tuning step, whose width is given by the resonator bandwidth, $\Delta f_{-3\, \mathrm{dB}}=f/Q_{\rm l}$. For the signal calculations at each frequency step, the frequency-dependent form factor and unloaded quality factor are derived from full electromagnetic simulations, as described in section~\ref{sec:cav_characterisation}. The use of the 3 cavity geometries DQLF-1 to -3 ensures continuous coverage over the frequency range from approximately 294 to 445~MHz. The nominal unloaded quality factor $Q_0(f)$ corresponds to a copper cavity operated at cryogenic temperature, as shown in ~\ref{sec:cav_characterisation} and plotted in Fig.~\ref{fig:UnloadedQ}.

We assume an overall M1 availability window of approximately 11 months. From this period, one month is conservatively allocated to the initial cooldown and preparation of the magnet. Two additional months are reserved for the two cavity-exchange operations required to cover the complete frequency range with DQLF-1, DQLF-2 and DQLF-3, including the associated shutdown, intervention and recommissioning time. This leaves eight months available for the physics campaign. Of this time, 10\% is reserved for rescanning candidate frequencies and performing additional validation measurements. Finally, we account for the data-taking live-time efficiency discussed in Sec.~\ref{sec:data-taking}. We conservatively adopt $\eta=0.82$, corresponding to the estimate obtained for the shortest dwell times considered in this proposal and therefore representing the least favourable DAQ duty cycle. The resulting effective exposure for the primary scan is thus
\[
t_{\rm live}
=
\left(11-1-2\right)\,\mathrm{months}
\times 0.90 \times 0.82
\simeq 5.9\,\mathrm{months},
\]
corresponding to approximately 180 days of accumulated axion-search data. Rather than assigning equal exposure to each cavity, this total live time is distributed uniformly among all frequency steps of the complete scan. Consequently, the exposure associated with each cavity is proportional to the number of tuning steps required to cover its frequency range, which depends on its frequency-dependent quality factor and resonant frequency. For the present cavity models, this corresponds approximately to 35\%, 38\%, and 27\% of the total live exposure for DQLF-1, DQLF-2 and DQLF-3, respectively, or about 62, 69 and 49 live days. The exact allocation used in the sensitivity calculation is obtained directly from the complete frequency-dependent cavity simulations.

The noise calculation is performed directly in terms of photon occupation
numbers rather than assuming a single frequency-independent system noise
temperature. Starting from the physical cavity temperature,
$T_{\rm phys}$, we calculate the corresponding effective cavity noise
temperature, including the Bose--Einstein thermal occupation and the
quantum vacuum contribution. The receiver noise is parametrized
independently by the number of added noise photons, $n_{\rm add}$, with
an equivalent added-noise temperature
$T_{\rm add}=n_{\rm add}h f/k_{\rm B}$. The total noise entering the
sensitivity calculation is then obtained from these contributions using
the cavity--receiver coupling model described in Sec.~\ref{sec:optimal_coupling},
and can equivalently be expressed as a frequency-dependent system noise
temperature $T_{\rm sys}(f)$. This treatment consistently accounts for
the physical cavity temperature, quantum corrections, receiver-added
noise, and the adopted cavity coupling and microwave receiver
configuration.

To account for the present uncertainties in detector performance, two
scenarios are considered. For the optimistic scenario, we assume
$T_{\rm phys}=30$~mK, corresponding to the target operating temperature
of the DQLF cryogenic system, and a receiver-added noise of
$n_{\rm add}=3.5$ photons referred to the receiver input. This level of
performance is representative of operation close to the quantum limit
and is consistent with the expected performance of the SQUID-based
receiver after optimization of the RF chain and implementation of
adequate cryogenic isolation, as discussed in
Sec.~\ref{sec:msa_characterization}. For reference, at the lower end of
the scan ($f \simeq294$~MHz), $n_{\rm add}=3.5$ corresponds to an
equivalent added-noise temperature of $T_{\rm add}\simeq50$~mK. In this
scenario, we assume operation without a cryogenic circulator between the
cavity and the first-stage amplifier. The cavity--receiver mismatch then
introduces a coupling-dependent cavity-noise contribution, which is
included explicitly in the sensitivity calculation. We operate at the
scan-rate-optimal coupling $\beta_{\rm opt}$ discussed in
Sec.~\ref{sec:optimal_coupling}, calculated at each tuning step from the
corresponding $Q_0(f)$ and noise parameters. The resulting frequency
dependence of $\beta_{\rm opt}$ is consistently propagated to the loaded
quality factor, signal extraction, noise, and scan step.

For the conservative scenario, we assume $T_{\rm phys}=100$~mK,
corresponding to the baseline cryogenic requirement, together with
$n_{\rm add}=8.5$ photons. At 290~MHz, the latter corresponds to an
equivalent added-noise temperature of approximately
$T_{\rm add}=120$~mK, a performance already experimentally demonstrated
with the present DQLF SQUID receiver even without the final cryogenic
isolation foreseen for the experiment. In this scenario, a cryogenic
circulator is assumed between the cavity and the receiver, providing a
matched microwave environment and therefore removing the
coupling-dependent reduction of the cavity-noise contribution. We
furthermore assume a conventional fixed coupling of $\beta=2$,
representative of robust operation without relying on full scan-rate
optimization. 

The two scenarios also bracket the expected cavity performance. The
optimistic scenario assumes a superconducting cavity coating capable of
increasing the unloaded cavity quality factor $Q_0$ by approximately one
order of magnitude with respect to the nominal cryogenic copper value,
whereas the conservative scenario assumes a quality factor five times
smaller than the nominal value, representative of the performance
observed in the first DQLF-1 prototype characterized in
Sec.~\ref{sec:kit_meas}. Together with the different assumptions on
physical temperature, receiver-added noise, microwave configuration,
and cavity coupling described above, these choices define the two
sensitivity projections presented below.

The number of frequency points required to cover the full scan range is determined by the loaded cavity bandwidth and therefore depends on both the unloaded quality factor and the cavity coupling, through
$Q_{\mathrm{l}}=Q_0/(1+\beta)$. Consequently, the number of scan points and the available integration time per tuning step differ substantially between the two scenarios. In the optimistic scenario, the higher cavity quality factor is partially compensated by operation at the scan-rate-optimized coupling, $\beta_{\rm opt}\simeq 8$--$11$, which broadens the loaded cavity bandwidth with respect to operation at lower coupling. In the conservative scenario, the lower cavity quality factor is combined with a fixed coupling $\beta=2$. For the assumed total exposure of 180 days, the resulting dwell times are typically at the level of a few hundred seconds in the optimistic scenario and extend to a few thousand seconds in the conservative case, with the shortest dwell times remaining at the $\sim100$~s level.

\begin{figure}[htb]
    \centering
    \includegraphics[width=0.7\linewidth]{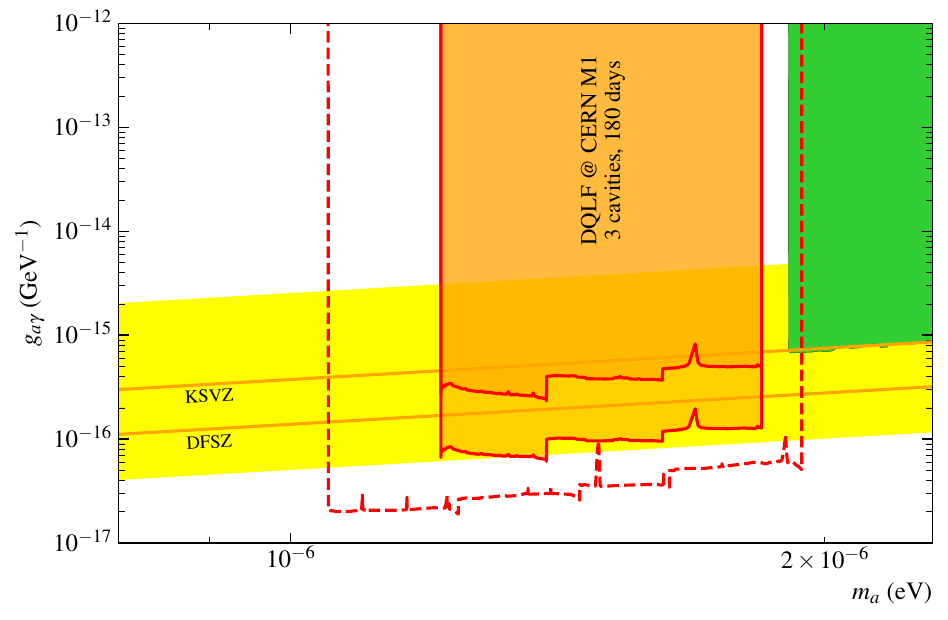}
\caption{
Projected sensitivity of the proposed DQLF search in the CERN M1 magnet to the axion--photon coupling $g_{a\gamma}$ as a function of the axion mass $m_a$. The solid red curves show the expected 90\% C.L. sensitivity for the three-cavity DQLF configuration and an effective exposure of 180 days, for the optimistic (lower curve) and conservative (upper curve) scenarios described in the text and summarized in Table~\ref{tab:sensitivity_assumptions}. The orange shading highlights the parameter space probed by the proposed search, with the two curves illustrating the range associated with the assumed experimental performance. The dashed red curve shows, for comparison, the projected sensitivity of the full-scale DQLF implementation in BabyIAXO under the assumptions of Ref.~\cite{Ahyoune:2023gfw} plus the improvements originally planned with DarkQuantum. Existing ADMX constraints are shown in green, while the yellow band indicates the QCD axion model region, and the orange lines the KSVZ and DFSZ benchmark models.
}
    \label{fig:rades_m1_sensitivity}
\end{figure}

The resulting projected sensitivity is shown in Fig.~\ref{fig:rades_m1_sensitivity}. Even under conservative assumptions, our proposal is expected to probe unexplored parameter space in the mass range around $1.2$\text{--}\SI{2}{\micro\electronvolt}. Under optimistic assumptions, the experiment approaches the lower edge of the parameter region predicted by benchmark QCD axion models, providing the first physics reach of the RADES/DQLF concept and establishing the technological basis for fully testing the DM axion paradigm in its subsequent implementation in BabyIAXO.

\begin{table}[t]
\centering
\caption{Summary of the assumptions adopted for the sensitivity projections.
The nominal cavity parameters are obtained from the electromagnetic simulations
described in Sec.~\ref{sec:cav_characterisation}. The two scenarios bracket
the expected performance in terms of cavity quality factor, physical
temperature, receiver-added noise, and microwave receiver configuration.
The full frequency dependence of the effective system noise is included in
the sensitivity calculation.}
\label{tab:sensitivity_assumptions}
\renewcommand{\arraystretch}{1.2}

\begin{tabular}{lcc}
\hline
Parameter & Optimistic & Conservative \\
\hline
Physical temperature $T_{\rm phys}$ & 30 mK & 100 mK \\
Receiver-added noise $n_{\rm add}$ & 3.5 photons & 8.5 photons \\
$T_{\rm add}$ at 290 MHz & $\sim$50 mK & $\sim$120 mK \\
Cryogenic circulator & No & Yes \\
Unloaded cavity quality factor & $10\times Q_0(f)$ & $0.2\times Q_0(f)$ \\
Coupling $\beta$ & $\beta_{\rm opt}(f)$ & 2 \\
Effective exposure & 180 days & 180 days \\
Confidence level & \multicolumn{2}{c}{90\% C.L.} \\
\hline
\end{tabular}
\end{table}

\section{Conclusions}
\label{sec:conclusions}

The search for axion dark matter is entering a decisive phase, with a rapidly growing international programme addressing complementary regions of the QCD axion parameter space. Despite these worldwide efforts, the 1\text{--}2 \SI{}{\micro\electronvolt} mass range remains essentially unexplored, primarily because it requires the combination of large magnetic volume, large resonant cavities, and ultra-low-noise microwave instrumentation. The proposed RADES-DQLF experiment directly addresses this experimental gap.

The combination of the CERN M1 superconducting magnet with the cryogenic, quantum-enhanced DQLF detector provides a unique opportunity to perform a competitive search for axion and axion-like particle dark matter in this unexplored region. Even under conservative assumptions regarding cavity quality factor and system noise temperature, the experiment reaches scientifically compelling sensitivity. Under the expected detector performance, it probes a substantial fraction of the QCD axion parameter space in the targeted mass range, while optimistic, yet realistic, operating conditions would allow sensitivity to benchmark models such as KSVZ and DFSZ over a large part of the explored frequency interval.

Importantly, the proposed experiment is not merely a technology demonstrator for the future BabyIAXO low-frequency programme. It constitutes a complete physics experiment with its own discovery potential and scientific impact. The deployment of three complementary cavity geometries allows continuous coverage of the targeted mass range, following the same modular strategy foreseen for the full BabyIAXO implementation. At the same time, the experiment naturally validates the detector technologies, cryogenic infrastructure, quantum-limited readout, and operational procedures that will later be employed in the full-scale DQLF experiment, thereby substantially reducing the technical risk associated with that future facility.

The experiment is also particularly timely. The availability of the M1 magnet during LS3 offers a unique opportunity to carry out a world-leading axion search substantially earlier than the intended final implementation in the BabyIAXO magnet. The collaboration is already supported by substantial resources through the ERC Synergy Grant \textit{DarkQuantum}, complemented by previous CERN PBC activities and by the long-standing experience of the RADES collaboration in designing, constructing and operating haloscope experiments at CERN. 

Beyond axion dark matter, the M1 setup may also offer opportunities for searches for other weak signals in the radio-frequency domain. In particular, the haloscope can exploit graviton-to-photon conversion in the external magnetic field through the inverse Gertsenshtein effect to search for high-frequency gravitational waves (HFGWs)~\cite{Aggarwal:2020olq}. Previous studies have investigated this possibility for the BabyIAXO haloscope configuration~\cite{Valero:2024ncz}, demonstrating the potential of large-volume resonant detectors operated in strong magnetic fields for HFGW searches. The corresponding sensitivity of the one-metre DQLF cavities is currently being investigated. The presence of several resonant TE modes could provide simultaneous sensitivity in different frequency intervals, while future optimization of the RF ports and receiver architecture could enable multi-mode and time-domain searches for transient HFGW signals.

In summary, we have laid out an experimental concept that combines a compelling standalone scientific programme with a clear strategic value for the future development of CERN's activities in axion searches and quantum-enabled detectors. By exploiting an existing  infrastructure at CERN, we aim to explore a completely uncharted region of parameter space with world-leading sensitivity.

\section*{Acknowledgments}

It is a pleasure to thank the CERN technical services along with several individuals for helping in the preparation of work.
In particular we would like to thank Gianluigi Arduini, Walter Wuensch, Nikolaos Charitonidis, Johannes Bernhard, Dragoslav Lasic, Bastian Rae, Guillaume Rosaz, Caroline Hain, Antti Kolehmainen and Jessica Golm.
We also would like to thank our colleagues from the MADMAX axion search experiment for discussion and coordination in preparation of the activities at CERN.

This work was supported by the European Research Council (ERC) under the European Union’s Horizon Europe programme (ERC Synergy Grant \textit{DarkQuantum}, grant agreement No. 101118911).

We acknowledge funding through the European Research Council under grant ERC-2018-StG-802836 (AxScale) as well as through the Deutsche For\-schungs\-ge\-mein\-schaft (DFG, German Research Foundation) under Germany’s Excellence Strategy – EXC 2094 – 39078331 as well as through Grant No. 532766533 (QUANTERA). 

This work also profited from strong support by CERN's PBC (Physics Beyond Collider) initiative.

This work was also supported by grants PID2022-137268NB-C51, PID2022-137268NB-C53, PID2025-171343NB-C53, funded by MCIN/AEI/10.13039/501100011033/ and by FEDER, “A way of making Europe”; and by the Plan Complementario de Astrofísica y Física de Altas Energías, funded by the European Union through NextGenerationEU within the Plan de Recuperación, Transformación y Resiliencia (PRTR), and co-funded by the Gobierno de Aragón.
We also acknowledge the Danish Innovation and Research Accelerator for CERN-collaboration (DIRAC), funded by the Danish Agency for Higher Education and Science (Grant No. 5313-00001B).

Funded by the European Union. Views and opinions expressed are however those of the author(s) only and do not necessarily reflect those of the European Union or the European Research Council Executive Agency. Neither the European Union nor the granting authority can be held responsible for them.

\bibliographystyle{utphys}
\bibliography{bib/axionbib}

\end{document}